\documentclass[twocolumn,usenatbib,useAMS]{mnras}
\PassOptionsToPackage{dvipsnames,svgnames,x11names}{xcolor}
\usepackage{xparse,soul}

\makeatletter
\NewDocumentCommand{\sotwo}{O{red}O{black}+m}
    {
        \begingroup
        \color{#1}%
        \setul{-.5ex}{.4pt}%
        \aef\SOUL@uleverysyllable{%
            \rlap{%
                \color{#2}\the\SOUL@syllable
                \SOUL@setkern\SOUL@charkern}%
            \SOUL@ulunderline{%
                \phantom{\the\SOUL@syllable}}%
        }%
        \ul{#3}%
        \endgroup
    }
\makeatother

\usepackage[flushleft]{threeparttable}

\usepackage{graphicx}	
\usepackage{amsmath}	
\usepackage{amssymb}	
\usepackage{multicol} 
\usepackage{multirow}
\usepackage{bm}	
\usepackage{pdflscape}
\usepackage{caption}
\usepackage{subfigure}
\usepackage{siunitx}
\usepackage{color}
\usepackage{lastpage}
\usepackage{xspace}

\newcommand{\kms}{\,{\rm km}\,{\rm s}^{-1}}
\newcommand{\kmsmpc}{\,{\rm km}\,{\rm s}^{-1}\,{\rm Mpc}^{-1}}

\newcommand{\simba}{\mbox{{\sc Simba}}\xspace}
\def\bsimba{\textbf{\textsc{Simba}}}
\newcommand{\hMpc}{\,h^{-1} \, {\rm cMpc}}
\newcommand{\hkpc}{\,h^{-1} \, {\rm ckpc}}
\usepackage[T1]{fontenc}
\usepackage{ae,aecompl}

\usepackage{newtxtext,newtxmath}
\newcolumntype{P}[1]{>{\centering\arraybackslash}p{#1}}
\newcolumntype{M}[1]{>{\centering\arraybackslash}m{#1}}

\def\citejap#1{\citeauthor{#1}\ \citeyear{#1}}

\title[CGM anisotropy in \simba]{Feedback-driven anisotropy in the circumgalactic medium for quenching galaxies in the \bsimba\ simulations}

\author[Yang et al.]{Tianyi Yang$^{1}$, Romeel Dav\'e$^{1,3}$, Weiguang Cui$^{1,2}$, Yan-Chuan Cai$^{1}$, John A. Peacock$^1$,
\newauthor and Daniele Sorini$^{4,1,5}$
\\
\\$^1$ Institute for Astronomy, University of Edinburgh, Royal Observatory, Blackford Hill, Edinburgh EH9 3HJ, UK
\\$^2$ Departamento de Física Teórica and CIAFF, Modulo 8 Universidad Autónoma de Madrid, 28049 Madrid, Spain.
\\$^{3}$ University of the Western Cape, Bellville, Cape Town 7535, South Africa
\\$^{4}$Institute for Computational Cosmology, Durham University, South Park Road, DH1 3LE, United Kingdom
\\$^{5}$D\'epartement de Physique Th\'eorique, Universit\'e de Gen\`eve, 24 quai Ernest Ansermet, 1211 Gen\`eve 4, Switzerland
}

\pubyear{2022}

\begin{document}
\label{firstpage}
\pagerange{\pageref{firstpage}--\pageref{lastpage}}
\maketitle

\begin{abstract}
We use the \simba galaxy formation simulation suite to explore anisotropies in the properties of circumgalactic gas that result from accretion and feedback processes.  We particularly focus on the impact of bipolar active galactic nuclei (AGN) jet feedback as implemented in \simba, which quenches galaxies and has a dramatic effect on large-scale gas properties.  We show that jet feedback at low redshifts is most common in the stellar mass range $(1-5)\times 10^{10}M_\odot$, so we focus on galaxies with active jets in this mass range. In comparison to runs without jet feedback, jets cause lower densities and higher temperatures along the galaxy minor axis (\simba jet direction) at radii $\ga 0.5r_{200c}-4r_{200c}$ and beyond.  This effect is less apparent at higher or lower stellar masses, and is strongest within green valley galaxies. The metallicity also shows strong anisotropy out to large scales, driven by star formation feedback.  We find substantially stronger anisotropy at $\la 0.5r_{200c}$, but this also exists in runs with no explicit feedback, suggesting that it is due to anisotropic accretion. Finally, we explore anisotropy in the bulk radial motion of the gas, finding that both star formation and AGN wind feedback contribute to pushing the gas outwards along the minor axis at $\la 1$~Mpc, but AGN jet feedback further causes bulk outflow along the minor axis out to several Mpc, which  drives quenching via gas starvation. These results provide observational signatures for the operation of AGN feedback in galaxy quenching.

\end{abstract}


\section{Introduction}\label{sec:intro}

The circumgalactic medium (CGM) is the gaseous environment surrounding a galaxy, which can extend up to hundreds of kpc from the galactic centre. The CGM is closely related to the process of galaxy evolution because it is the site where galactic inflows and outflows interplay \citep[see e.g.][and references therein]{CGM_review}. Cold gas in the CGM accretes onto the central galaxy, fuelling future star-forming activity. Meanwhile, gas can be carried out into the CGM by galactic-scale outflows, which causes gas depletion in the central regions and results in suppression of star formation there. These outflows can be driven by stellar feedback in low-mass galaxies, feedback from active galactic nuclei (AGN) in high-mass galaxies, or even a combination of the two mechanisms \citep[see e.g.][and references therein]{Somerville_Dave_review}.  

AGN feedback effects, which are sourced by gas accretion onto supermassive black holes (SMBH), are known to be a major source of energy input to the CGM. AGN feedback is thought to be an important ingredient in regulating the growth of central black holes and suppressing the star-forming activity in galaxies \citep[e.g.][]{Guillard_AGN_3C,Morganti_AGN_2017,Harrison_SMBH_AGN_sfr}. According to observations, SMBH exist at the centres of most massive galaxies \citep[see e.g.][and references therein]{SMBH_ref}, and the AGN feedback mechanisms appear in two modes: `quasar' and `radio' mode \citep[see e.g.][]{Ho_review,Fabian_2012, AGN_summary}. The `quasar' mode is found in luminous AGN with high accretion rates. In this case, the feedback energy is released into the surroundings in the form of radiation from the central accretion disk, which can drive powerful winds. The `radio' mode is usually found in galaxies hosting AGN with low accretion rates. These are radiatively inefficient but are capable of releasing large amounts of feedback energy into the CGM by means of bubbles or radio jets. Relatively clear observational evidence exists for AGN feedback in action (e.g. \citejap{2012MNRAS.425L..66M}), and results from \simba\ simulations have suggested that this mechanism is key to the quenching of galaxies \citep{Cui2021}. Nevertheless, much remains to be understood about the detailed operation of these processes, and in particular how they interact with the CGM.

Hydrodynamical simulations provide opportunities for studying how AGN feedback reshapes the properties and evolution of the CGM. To reproduce various observed galaxy properties, it is necessary to include the modelling of AGN feedback in simulations: this enables suppression of star formation in massive galaxies and prevents them undergoing excessive growth \citep{Cosmo_owls_le_brun, EAGLE_ref, Sijacki_illustris_coeval_BH, McCarthy_BAHAMAS,TNG_feedback_Weinberger,cui2018,simba_ref}. AGN feedback is also important in allowing simulations to reproduce other observed thermodynamic and chemical properties of the gas, such as the hot gas fraction in groups and CGM metal absorption line properties \citep{McCarthy_2010, simba_ref, Oppenheimer_2021_review, Cui2022}.

Although simulations have successfully reproduced a wide range of observed galaxy properties, the implementations of SMBH feedback in different codes are distinct in a number of ways. Regarding the form of feedback energy, there are two major ways of implementing this: either by heating up the surrounding gas isotropically with the expected amount of energy from the AGN \citep[e.g. in cosmo-OWLS and EAGLE simulations:][]{Cosmo_owls_le_brun,EAGLE_ref}, or by ejecting gas particles with kinetic kicks along a random or bipolar direction \citep[e.g. in IllustrisTNG and SIMBA simulations:][]{TNG_feedback_Weinberger, simba_ref}. Both mechanisms succeed in matching galaxy observations, but the way in which the released energy propagates into the CGM must be different. This may have a strong impact on the properties of the CGM and thus produce distinctive observable features in the resulting CGM gas distribution.

The interplay between galactic feedback outflows and the CGM has been widely studied in both observations and simulations, over a large range of redshifts and stellar masses. This includes, for example: the distribution of highly ionised gas such as oxygen \citep[e.g.][]{Oxygen_TNG, OIII_emission}; the abundance of Mg\textsc{II}-traced cold gas \citep[e.g.][]{Bordoloi_MgII_bipolar, Kacprzak_bimodal_MgII, Bouch_MgII_bipolar, Nielson_MgII_bipolar_accretion, MgII_emission_TNG}; emission lines such as 21\,cm and H$\alpha$ \citep[e.g.][e.g.]{HI_21_cm, H_alpha_emission_CGM}; the warm diffuse gas via the thermal Sunyaev-Zeldovich (tSZ) effect \citep[e.g.][]{Lokken_threehundred_cluster_SZ,Tianyi_tSZ,SZ_cavity}; $\mathrm{Ly}\alpha$ and metal absorption lines \citep[e.g.][]{turner_2014, meiksin_2014, meiksin_2015, meiksin_2017, turner_2017, sorini_2018, sorini_2020, appleby_2021, appleby_2023}; and the hot dense atmosphere probed by X-rays \citep[e.g.][]{TNG_X_ray_CGM,X_ray_TNG_CGM2, TNG_CGM_anisotropy}. In particular, some observations have shown that, for disk-dominated galaxies, galactic outflows emerging from the disk are preferentially ejected biconically into the CGM, where the strongest outflow features are captured along the minor axis of the disk \citep[e.g. in][]{Bordoloi_MgII_bipolar, Bouch_MgII_bipolar}. Infalling gas is preferentially accreted in the galaxy plane \citep[e.g. in ][]{Bouche_accreting_cold_gas, Nielson_MgII_bipolar_accretion}. For red passive galaxies, however, their CGM distribution tends to be more isotropic \citep[e.g. in][]{Kacprzak_bimodal_MgII, Nielson_MgII_bipolar_accretion}. In simulations, although different feedback models are implemented, this type of angular dependence is also widely found \citep[e.g.][]{peroux_metal_CGM_TNG,Eagle_outflow_Mitchell, TNG_x_ray_bubble}. These studies suggest that the CGM properties connect closely to the feedback activity inside galaxies. Specifically, outflow features in the CGM are generally more prominent along the minor axis of the disk (edge-on projection), and accretion is more easily observed along the galaxy disk (face-on projection).

However, the detailed outflow features and their angular dependence are quite sensitive to the adopted AGN feedback model, which in turn affects the predicted CGM distribution and the galaxy evolution processes. In particular, the outflow angular dependence and the predicted CGM properties can differ significantly in the EAGLE and TNG simulations \citep{TNG50_galactic_outflows_driven_by_supernovae_and_black_hole_feedback, Eagle_outflow_Mitchell,EAGLE_TNG_comparison}. Under the TNG framework, \citet{Pillepich_TNG_illustris} and \citet{TNG_model_variants} found that the resulting $M_{\rm BH}-M_{\ast}-$sSFR relation of galaxies depends strongly on the chosen parameters of the AGN feedback model in TNG. They further confirmed that kinetic wind feedback is required in order to reproduce a quiescent galaxy population that is consistent with observations. Furthermore, the CGM anisotropic features can be altered by the form of released energy, depending on whether the kinetic or the thermal feedback mode dominates around galaxies \citep{Zinger_TNG,CGM_Milky_Way_like_ani}. The sensitivity of CGM anisotropy to AGN models has been further discussed by comparing results predicted from simulations to observations, such as the X-ray hardness \citep{TNG_CGM_anisotropy} and the satellite distribution around their centrals \citep{quenched_fraction_anisotropy}. Therefore, by studying the feedback-driven anisotropy in the CGM, we aim to identify and predict the observational consequences owing to feedback. This provides possible ways to constrain AGN models with further observations, which is a crucial step towards understand how galaxies evolve and undergo quenching.

In this paper, we focus on the anisotropic behaviour of the CGM and its relation to jet activity using \simba simulations, including in particular different runs with various feedback models turned on/off. We study the spatial CGM distribution around central galaxies by stacking a large sample of simulated galaxies. The CGM anisotropy is quantified by the quadrupole
moments of various physical quantities, and we further examine the dependence of these signals on the implemented AGN models, central galaxy mass and star formation status. Finally, we explore the anisotropy in gas radial motion and the redshift evolution of the feedback-driven anisotropic features. These provide us insight into how different feedback mechanisms regulate galactic outflow and eventually drive galaxy quenching under the \simba framework. 

The paper is organised as follows. We first introduce a brief summary of the \simba simulation suits, especially the implemented feedback models, as well as our methodology in \S\ref{sec:method}. Then in \S\ref{sec:results}, we present our main anisotropic results regarding the CGM properties considered in this work: mass, density, metallicity and thermal pressure. Around \simba-100 central galaxies, we study how the angular dependence of CGM distribution varies with their host properties, such as star formation status (\S\ref{ssec:galaxy_type_anisotropy}) and mass (\S\ref{ssec:angular_mass_anisotropy}). Then in \S\ref{ssec:model_anisotropy}, we present the effect of feedback models on the resulting CGM properties, using \simba-50 model variants. In \S\ref{sec:Impact_of_feedback_activity_on_galaxy_quenching}, by exploring the bulk radial gas motion around \simba-50 galaxies (\S\ref{ssec:phase-space_diagram}) and by tracing progenitors of $z = 0.0$ quenched samples (\S\ref{ssec:Redshift_evolution_of_feedback}), we discuss the connection of AGN feedback models in \simba-50 variants and further see what drives galaxy quenching in \simba model. We discuss and summarise our main findings in \S\ref{sec:discusion}, \S\ref{sec:discussion_2} and \S\ref{sec:conclusions}.

\section{Methodology}\label{sec:method}

We begin with an overview of the \simba simulations in \S\ref{ssec:sim_sec}. Owing to its importance for this work, we summarise the modelling of black hole feedback adopted in \simba models in \S\ref{ssec:simba_feedback}. We move on to the selection of galaxies in \S\ref{ssec:galaxy_selection}, and finally introduce our methodology used to characterise the anisotropy of CGM properties in \S\ref{ssec:map_stacking} and \S\ref{ssec:anisotropy_characterize}. 

\subsection{The \bsimba ~simulations}\label{ssec:sim_sec}
\begin{table}
  \centering
  \begin{threeparttable} 
    \caption{Feedback descriptions for different \simba-50 runs. Note that the Simba-100 fiducial run employs the same `allphs' model as the \simba-50 run.}
     \label{tab:simba_variant_table}
  \begin{tabular}{|M{1.5cm}|M{1.0cm}|M{1.0cm}|M{1.0 cm}|M{1.0cm}|}
    \hline
    Model & stellar feedback        & AGN radiative mode &AGN jet mode\tnote{a}& X-ray heating \\ \hline
    `allphys'& \checkmark & \checkmark&\checkmark, ~~~~with $f_{\rm Edd}$\tnote{b} and $M_{\rm BH,lim}$ cut\tnote{c} &\checkmark, ~coupled with jet  \\ \hline
    `nox'&\checkmark &\checkmark&\checkmark& \text{\sffamily X}  \\ \hline
    `nojet'& \checkmark & \checkmark &\text{\sffamily X}&\text{\sffamily X}  \\ \hline
    `noagn'& \checkmark & \text{\sffamily X} &\text{\sffamily X}&\text{\sffamily X}  \\ \hline
    `nofb'& \text{\sffamily X} & \text{\sffamily X} &\text{\sffamily X}&\text{\sffamily X}  \\ \hline
  \end{tabular}
  
    \begin{tablenotes}
    \item[a] The criteria for the jet AGN feedback to be turned on are $f_{\rm Edd}$ [b] and $M_{\rm BH}$ [c]. Full jet speed is achieved when $f_{\rm Edd} <0.02$, see \autoref{eqn::AGN_wind_jet}. 
    \item[b] $f_{\rm Edd} <0.2$ 
    \item[c] $M_{\rm BH} \geq M_{\rm BH,lim} = 10^{7.5} M_{\odot}$
  \end{tablenotes}
\end{threeparttable}
\end{table}

\simba \citep{simba_ref} is a suite of hydrodynamic simulation using the $\textsc{Gizmo}$ code \citep{Gizmo_ref}. Dark matter and gas particles are evolved within a periodic cubical volume with a cosmology broadly concordant with $\textit{Planck}$ 2015  \citep{planck_2015_cosmo_paper}: $\Omega_{m,0} = 0.3, \Omega_{\Lambda,0} = 0.7, \Omega_{b,0} = 0.048, H_{0} = 68\kmsmpc, \sigma_{8} = 0.82$ and $n_{s} = 0.97$. The fiducial run (denoted \simba-100) has a box length of 100 \textrm{comoving $h^{-1}$\,Mpc} (hereafter $h^{-1}$\,cMpc), evolving from $z = 249$ to $z = 0$ with $1024^{3}$ dark matter particles and $1024^{3}$ gas elements. To explore the variation of anisotropic features with input feedback models, there are several $50\hMpc$ boxes (denoted \simba-50) using $512^{3}$ dark matter particles and $512^{3}$ gas elements. The mass resolution for both cases is $1.82\times10^{7} M_{\odot}$ for gas cells and $9.58\times10^{7} M_{\odot}$ for dark matter particles. The initial conditions for all \simba runs are identical for a given box size.

In \simba, star formation is modelled using an $\rm H_{2}-$based star formation rate. This is given by the $\rm H_{2}$ density divided by the dynamical time with $\textrm{SFR} = \epsilon_{\ast} \rho_{\rm H_{2}}/t_{\rm dyn}$, where $\epsilon_{\ast} = 0.02$ \citep{H2_sfr_eq}. The $\rm H_{2}$ fraction is calculated using the subgrid model of \citet{SFR_model} based on the metallicity and local column density, with some minor modifications to account for the variations in numerical resolution \citep{MUFASA_ref}. The chemical enrichment model tracks eleven elements in total (H, He, C, N, O, Ne, Mg, Si, S, Ca, Fe) from Type II supernovae (SNe), Type Ia SNe, and Asymptotic Giant Branch (AGB) stars. The star formation-driven galactic winds are modelled as decoupled two-phase metal-loaded winds, with 30$\%$ of ejected wind particles being hot and with a redshift-independent mass loading factor that scales with stellar mass \citep{mass_loading_factor_scaling}. The SF wind velocity, modified from the scaling in \citet{wind_vel_ref}, is computed via the following equation:
\begin{equation}\label{eqn::sf_wind_vel}
    v_{w} = 1.6 \left(\frac{v_c}{200\kms}\right)^{0.12} v_{c} + \Delta v(0.25 R_{\rm vir}),
\end{equation}
where $v_{c}$ is the galaxy's circular velocity at $0.25R_{\rm vir}$, and $\Delta v(0.25 R_{\rm vir})$ is an extra velocity kick corresponding to  the gravitational potential difference between the wind launch radius and $0.25R_{\rm vir}$ \citep{wind_vel_ref_2}.

Black holes are seeded and grown during the simulation, and the accretion energy drives feedback that causes star formation to become quenched. Black hole growth in \simba is modelled with a two-mode accretion model. For cold gas with $\rm T<10^{5} K$, the gas inflow is implemented using the torque-limited accretion model \citep{cold_gas_accretion_model}. While hot gas ($\rm T>10^{5} K$) is accreted onto black holes via Bondi accretion \citep{Bondi_accretion}. The major improvement of the black-hole growth model adopted by \simba is the torque-limited accretion for the cold gas, which does not require the black hole to self-regulate its own growth \citep{BH_growth_ref}. This allows for the implementation of a more physical AGN feedback model, which will be discussed in \S\ref{ssec:simba_feedback}. Other input physical mechanisms such as radiative cooling and heating, the formation and evolution of dust are also included in \simba runs. Specifics of these models are available in \citet{simba_ref}.

\subsection{Black hole feedback models in \bsimba}\label{ssec:simba_feedback}

AGN feedback is responsible for quenching galaxies in \simba. It is implemented by a two-mode model, which is motivated by the observed dichotomy in black hole growth \citep[e.g. in ][]{AGN_summary}. A `radiative mode' is applied when a black hole is accreting at high Eddington ratios ($f_{\rm Edd} = \dot{M}_{\rm BH}/\dot{M}_{\rm Edd}$), which mimics the molecular and warm ionised gas outflow \citep{radiative_mode_ref1}. AGN wind particles are ejected without modifications of the gas temperature, with a typical electron temperature of $\sim 10^{4}\,\rm K$ and with a velocity of $\sim\,$$ 1000\kms$ \citep{radiative_mode_ref2}. In this case, the outflow velocity is related to the black hole mass via:
\begin{equation}\label{eqn::AGN_wind_rad}
    v_{w,\rm EL} = 500+500(\textrm{log}_{10}M_{\rm BH} - 6)/3 \kms.
\end{equation}
While at low Eddington ratios, a `jet mode' is applied to drive high-velocity hot gas outflows \citep{Fabian_2012}. The jet direction is set parallel/anti-parallel to the inner gas disk of the host galaxy. Its angular momentum vector is computed using the 256 closest gas particles around the central black hole (typically $\sim\,$1\,kpc), with an upper limit of $R_{\rm inner ~disk} \leq2\hkpc$. Gas particles are heated to the virial temperature of the halo before ejection and the velocity becomes stronger as $f_{\rm Edd}$ drops:
\begin{equation}\label{eqn::AGN_wind_jet}
    v_{w,\rm jet} = v_{w,\rm EL}+7000\,\textrm{log}_{10}(0.2/f_{\rm Edd})\; \kms.
\end{equation}
The transition between radiative and jet mode happens when black holes have $f_{\rm Edd} < 0.2$ and $M_{\rm lim, BH} \geq 10^{7.5} M_{\odot}$. The velocity increase is capped at $7000 \kms$ above $v_{w,\rm El}$ when  $f_{\rm Edd} \leq 0.02$.

Finally, X-ray heating by the accretion disk is also included by \simba when its jet model is turned on and gas fractions within the black hole kernel are below 0.2, as motivated by \cite{Choi_2012}. This mimics the deposition of high-energy photons into the surrounding gas and is implemented in two modes: for non-ISM gas (with hydrogen number density of $n_{\rm H} < 0.13\,\rm cm^{-3}$), gas temperature values are directly increased based on the local radiation flux. For ISM gas, half of the radiation energy is applied as a radial kick outwards to gas particles, while the remainder is added as heat. X-ray feedback causes only modest changes to the galaxy stellar mass function, but it is crucial in order to achieve full quenching of the star formation in massive galaxies \citep[see \S4 in][]{simba_ref}.

There are several model variants implemented in the \simba-50 run: `allphys', which includes all the aforementioned physics identical to the \simba-100 fiducial run; `nojet', which turns off the bipolar jet and X-ray feedback; `noagn', which only includes stellar feedback but with all AGN feedback turned off; `nox', which only turns off the X-ray feedback; and `nofb', where all explicit feedback is turned off. A brief description and summary of these models is given in Table \ref{tab:simba_variant_table}. In \S\ref{ssec:model_anisotropy}, we show a comparison of our results for various different \simba-50 runs when discussing the sensitivity of CGM properties to the physics models. After convergence tests between \simba-50-`allphys' and the \simba-100 run (Appendix \ref{sec::convergence_test_full_physics_models}), we thereafter include results from \simba-100 only, where more samples are available for further analysis.

\subsection{Galaxy selection}\label{ssec:galaxy_selection}

\begin{figure*}
\centering
    \includegraphics[width=0.8\linewidth]{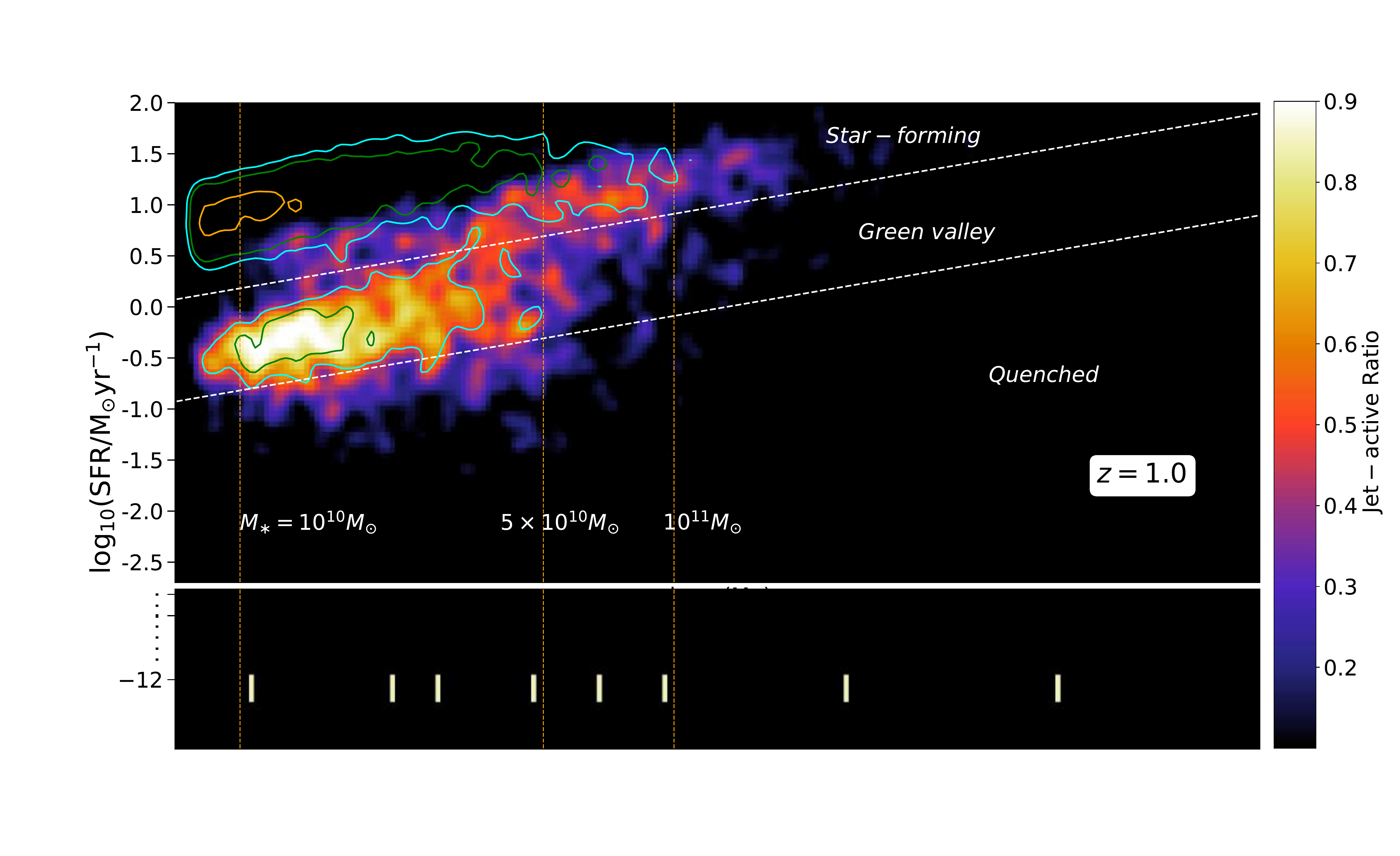}\\
    \vspace{-2.0 cm}
    \centering
    \includegraphics[width=0.8\linewidth]{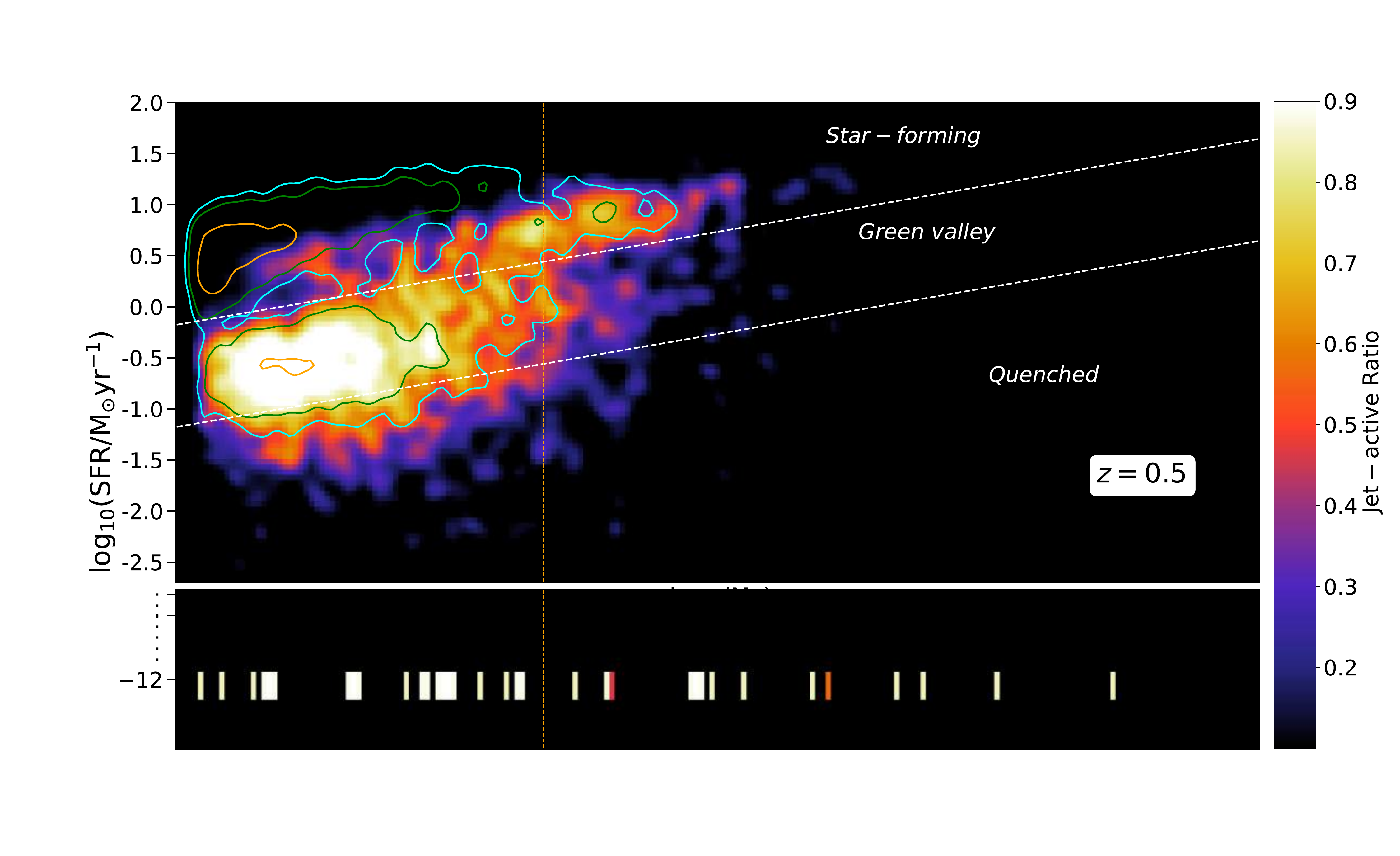}\\
    \vspace{-2.0 cm}
    \centering
    \includegraphics[width=0.8\linewidth]{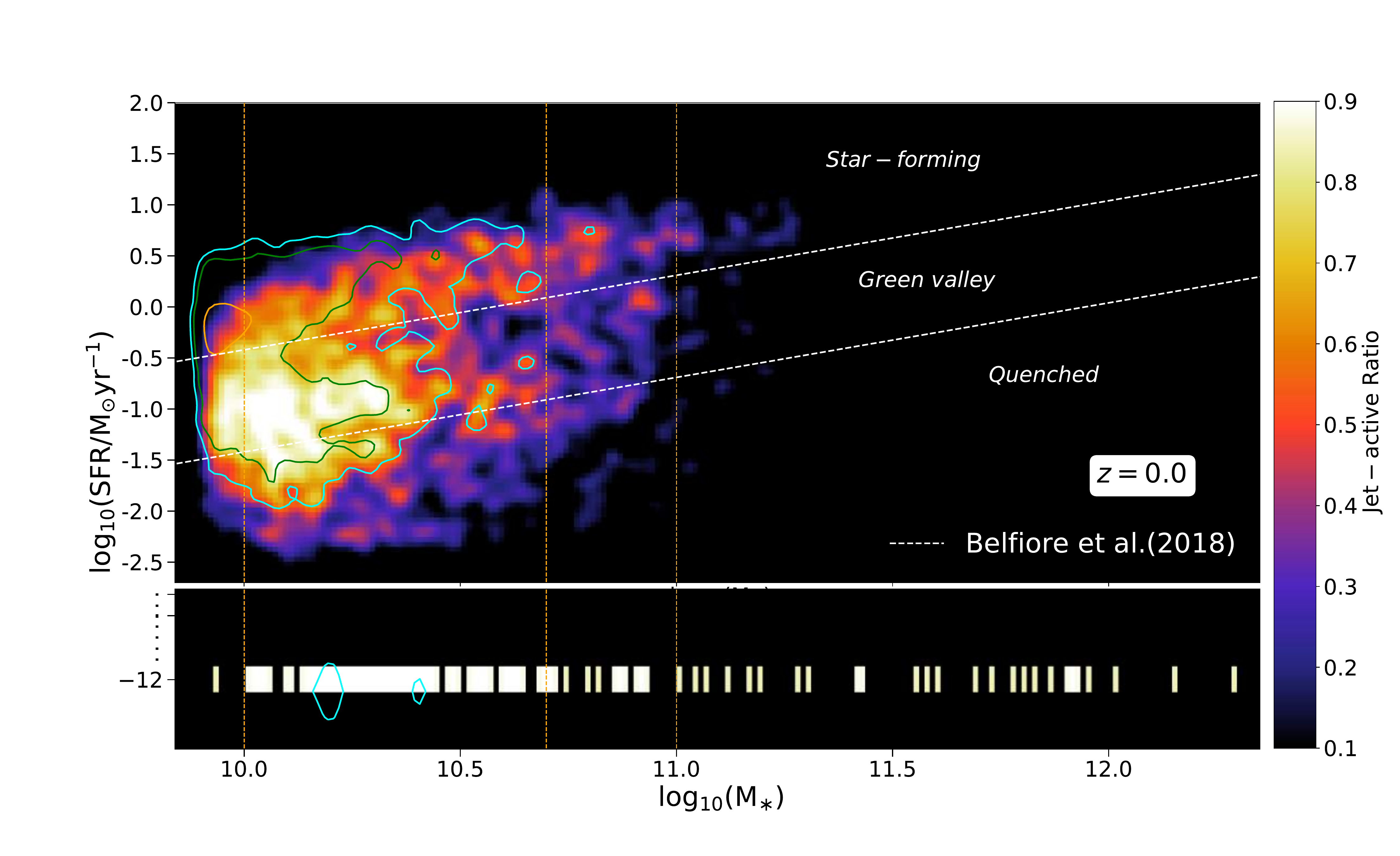}
    \vspace{-0.5 cm}
\caption{Star formation rate vs stellar mass relation for all central galaxies (the main sample) in the \simba-100 run at $z$ = 1.0 (\textit{top}), 0.5 (\textit{middle}) and 0.0 (\textit{bottom}). The colourbar shows the jet-active ratio at different redshifts (as defined in \S\ref{ssec:galaxy_selection}). Contours present the number distribution of main sample galaxies with $M_{\ast}\geq 10^{9.9}$. This lower $M_{\ast}$ cut is empirically chosen so that the contours overlap with the colourmap. Regions of star forming (SF), green valley (GV) and quenched (Q) galaxies are demarcated by the white dashed lines, which are the selection criteria adopted by \citet{Belfiore_SFMS} (see text for details). Vertical dashed lines divide the galaxies into three stellar mass bins that will be considered in further analysis below. It is noticeable that at all redshifts, `jet-active' galaxies are the most populous within the GV region between $M_{\ast} = 10^{10} M_{\odot}$ and $5\times10^{10} M_{\odot}$.}
\label{central_galaxies_dist}
\end{figure*}

The main sample considered in this study consists of the galaxies at $z = 0$ that are central, have black hole accretion rate $> 0$ and stellar mass $>10^{10} M_{\odot}$. In this work, stellar mass (denoted by $M_{\ast}$) is defined as the total mass of stellar particles within 30 $\rm ckpc$ spherical apertures. The choice of this stellar mass cut is adopted owing to the AGN fraction$-M_{\ast}$ plot from the 
simulations (Figure \ref{central_galaxies_dist}, which is discussed later) as well as from observations \citep[e.g.][]{Kauffmann_2003_AGN}, which suggest that the strong AGN fraction declines significantly below $M_{\ast}=10^{10} M_{\odot}$. We also divide our main sample into the following sub-catalogues when necessary:

\begin{itemize}
    \item \textit{`Jet-active' galaxies}. For a better visualisation of the anisotropic features, we also analyse galaxies with low Eddington accretion ratio and high central black hole mass ($0<f_{\rm edd}<0.02$, $M_{\rm BH}\geq10^{7.5} M_{\odot}$). As discussed in \S\ref{ssec:simba_feedback}, these are the criteria adopted in the `allphys' model when full jet speeds are achieved. Note that we will use the term `jet-active galaxies' to refer exclusively to galaxies that have jets with the full jet speeds. This is only meaningful in the `allphys' or `nox' models when a jet is implemented. For other models that have no jets, but comparing with simulations run with the same initial conditions as in the `allphys' model, we can identify galaxies that are the counterparts of those in the `allphys' simulations. We will refer to these galaxies as `jet-active counterparts'.
    
    \item \textit{Galaxy type}. Galaxies are categorised into star forming (SF), green valley (GV) and quenched (Q) galaxies based on the observed star-forming main sequence (SFMS) in \citet{Belfiore_SFMS}. Their best fit line is given by:
    \begin{equation}\label{eqn::belfiore_SFMS}
        \textrm{log}(\textrm{SFR}/M_{\odot} \textrm{yr}^{-1})= 0.73 ~\textrm{log}(M_{\ast}/M_{\odot})-7.33,
    \end{equation}
    with a scatter of 0.39 dex. According to this line, the lower boundary of SF galaxies is the upper dashed white line shown in Figure \ref{central_galaxies_dist}, which is $1\sigma$ below the best-fit SFMS line. GV galaxies are defined as having SFR values down to 1 dex below this line. Quenched galaxies are therefore all samples below the GV region. To account for the redshift evolution of the main sequence, we empirically boost their normalisation by a factor of $(1+z)^{2}$ when considering $z>0$.
\end{itemize}

Figure \ref{central_galaxies_dist} shows the SFR$-M_{\ast}$ plots for all central galaxies in \simba-100 at three different redshifts ($z = 1.0$, 0.5 and 0.0), overplotted with the jet-active ratio and number contours of main sample galaxies. Jet-active ratio is defined as the ratio between the numbers of galaxies in the `jet-active' sample and in the main sample, per SFR$-M_{\ast}$ bin. Galaxies are demarcated as SF, GV and Q based on the boundary lines discussed above. Vertical dashed lines group the galaxies into three stellar mass bins that will be considered in this study. We will show in this study that AGN jet feedback causes more galaxies to enter into the GV region as time evolves, and these eventually become quenched. Furthermore, at all redshifts the majority of `jet-active' galaxies reside in the GV region with $10^{10} M_{\odot}<M_*<5\times10^{10} M_{\odot}$, suggesting that the efficiency of jet production peaks for these galaxies. This is also broadly consistent with observations \citep[e.g. in][]{Nandra_AGN_obs,Schawinski_AGN_obs,Povic_AGN_obs}.

Additionally, we further include an analysis regarding how the anisotropic features evolved with redshift. We will discuss the sample selection of this analysis below.

\subsection{Gas properties around galaxies via stacking}\label{ssec:map_stacking}

\begin{figure*}
    \begin{minipage}[b]{1.0\textwidth}
        \centering
        \includegraphics[width=\linewidth]{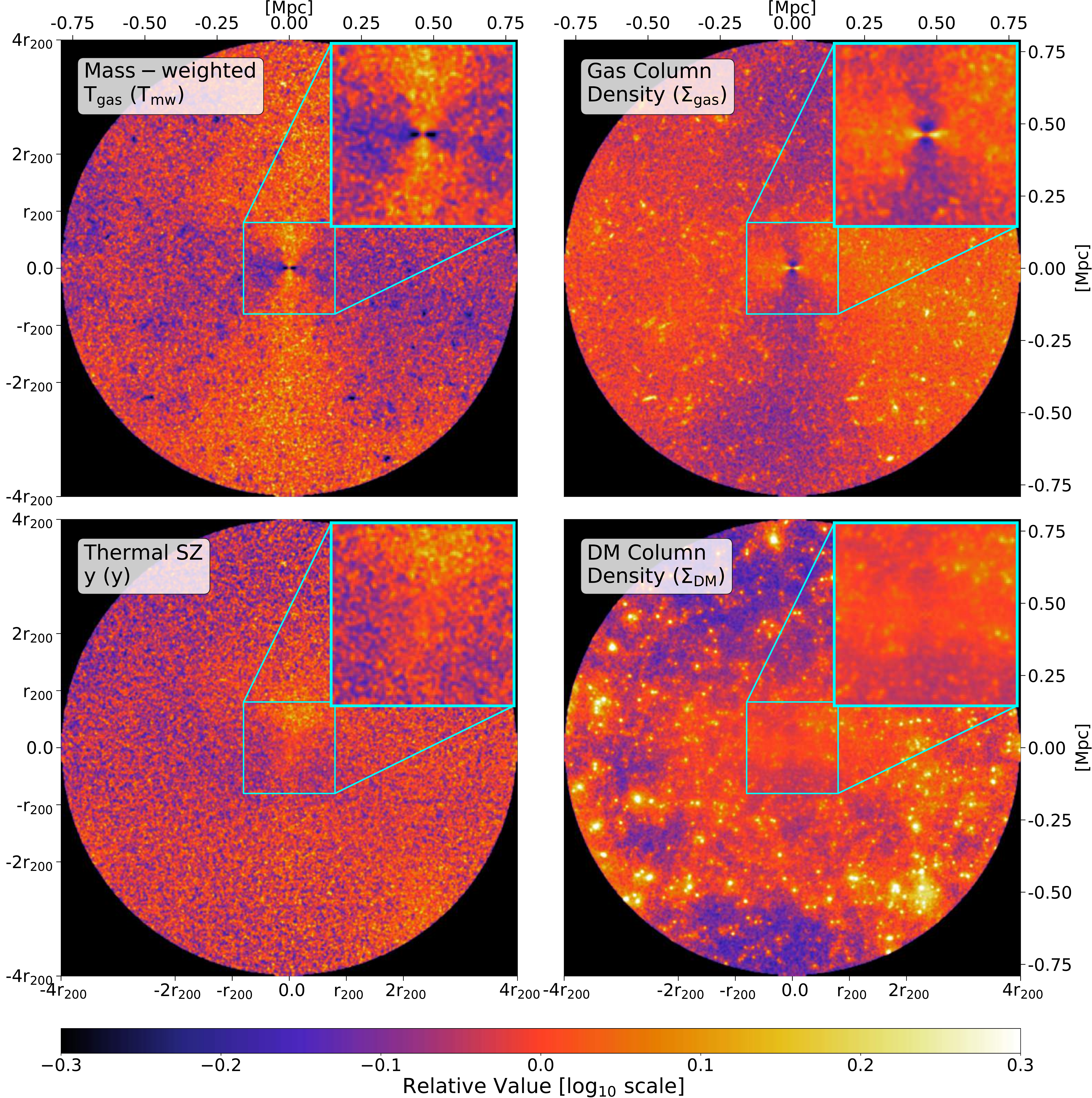}
    \end{minipage}%
    \vspace{+0.1 cm}
\caption{Anisotropy of the CGM properties for `jet-active' central galaxies at $z$ = 0.0 (edge-on view). Galaxies are selected from the \simba-50-`allphys' model with stellar mass between $1\times10^{10} ~M_{\odot}$ and $5\times10^{10} ~M_{\odot}$. There are 377 galaxies in total and the averaged host halo mass is $\langle \textrm{log}_{10} (M_{200c}) \rangle = 12.04$. \textit{Top row}: mass-weighted temperature map ($T_{\rm mw}$, \textit{left}) and gas column density map ($\Sigma_{\rm gas}$, \textit{right}). \textit{Bottom row}: thermal SZ $y$ map (SZ$-y$, \textit{left}) and dark matter column density map ($\Sigma_{\rm dm}$, \textit{right}). Maps are are normalised with respect to the azimuthal average. Minor axes of galaxies are aligned along the inner disk momentum directions (jet directions in this case), as well as all particles within a cubic region of $\pm4r_{200c}$. Note that the topmost and leftmost coordinates are in units of $\rm cMpc$ for comparison. On each panel, the small window in the upper right corner zooms into the central square region with size $r_{200c}$.}
\label{SIMBA50_allphys_bh_sfr_1e10_5e10_s151}
\end{figure*}

Since the results obtained from individual galaxies can be noisy, maps of gas properties are combined together via stacking. Mean galaxy properties are evaluated in either edge-on or face-on projections. These are then rotated with respect to the minor axes of their inner gas discs -- which is the assumed jet direction (see the definition in \S\ref{ssec:simba_feedback}), as well as using all gas particles within a cube of size $4\times r_{200c}$. Here, $r_{200c}$ is our adopted definition of the virial radius: the radius within which the mean density is 200 times the critical density. Therefore, edge-on projection is when the vertical direction is aligned along the minor axes of the galaxies, while the face-on projection is when viewing the galaxies along the plane of their `inner discs'. We deliberately chose to stack galaxies along this direction because this is the bipolar feedback direction as defined in the \simba-`allphys' model, where one should in principle `observe' the most anisotropic features in the stacked map if there are any. For comparison, we also repeat the same analysis by stacking galaxies along their stellar angular momentum directions, as discussed in Appendix \ref{sec::convergence_test_stacking}. This does not affect the anisotropic features of the CGM properties.

For the projection, particle column densities are derived by summing along the line of sight, while averaged quantities such as gas temperature and metallicity are computed in a mass-weighted manner along the line of sight. Gas metallicity is converted into solar units by dividing by the solar metal mass fraction, for which we use 0.0127. Only non star-forming gas cells are selected while analyzing CGM gas properties. To highlight the spatial anisotropy of the CGM properties, we further normalise the stacked map with respect to the azimuthal average. All maps have a resolution of $500\times500$ pixels. The coordinates are standardised such that all galaxies have the same size in units of the virial radius while stacking. The resolution in physical units corresponds to $\sim\,$3.2\,$\rm ckpc$ at the lowest stellar mass bin (with $M_{\ast} = 1\times10^{10}-5\times10^{10} M_{\odot}$) and $\sim\,$6.5\,$\rm ckpc$ at the highest mass bin (with $M_{\ast} > 1\times10^{11} M_{\odot}$).

Figure \ref{SIMBA50_allphys_bh_sfr_1e10_5e10_s151} shows the edge-on maps of CGM properties around central `jet-active' galaxies selected from \simba-50-`allphys' at $z = 0$, for the lowest stellar mass bin. We present the gas temperature $T_{\rm mw}$, gas column density $\Sigma_{\rm gas}$, expected SZ$-y$ signal and dark matter column density $\Sigma_{\rm DM}$ on the map. From these maps, it is obvious that the CGM distribution is bipolar, with the direction of the pole being aligned with the direction of the jet.  

\subsection{Characterizing the anisotropy}\label{ssec:anisotropy_characterize}

In our study, the angular location of the CGM is defined with respect to the minor axis of inner gas disk (\S\ref{ssec:map_stacking}). Therefore, an azimuthal angle of $0^{\circ}$ denotes alignment with the minor axis, where for models with AGN feedback turned on this is identical to the outflow direction. We first average the four quadrants of the stacked maps (e.g. Figure \ref{SIMBA50_allphys_bh_sfr_1e10_5e10_s151}), assuming that no distinctive features are present in any particular quadrant. Then, to characterise the angular dependence of the CGM properties, we compute the quadrupole moment of CGM properties in the stacked map as follows:
\begin{equation}\label{eq::quad_cal}
    \xi_{\ell}^{r}(r) = \int_{0}^{1} \xi(r, \mu)(1+2\ell)P_{\ell}(\mu)  \,d\mu, 
\end{equation}
where $\xi(r, \mu)$  is some property of the CGM; $\ell=2$; $\mu = \textrm{cos}\theta = y_{\rm corr}/r$; $\mu = 1$ corresponds to to angular alignment with the minor axis; $r$ is the 2D projected distance from the galactocentric centre; and $P_{\ell} (\mu)$ are the Legendre polynomials with $P_{2} (\mu) = \frac{1}{2}(3\mu^2-1)$. If the CGM distribution $\xi(r, \mu)$ is purely isotropic around galaxies, the above integration yields zero. Otherwise, it yields a positive value if the CGM tends to be distributed closer to the minor axis, or a negative value if the distribution is along the galaxy plane (a disk-like feature). We will denote these measured quadrupole moments with a subscript according to the quantity of interest: $\xi_T$ for the projected CGM temperature etc.

\section{Anisotropic distribution of CGM in simulations}\label{sec:results}
We expect the CGM properties to be regulated by AGN feedback processes, which is also connected to the host galaxies of the AGN, and possibly also their host dark matter haloes.
In this section, we investigate how the CGM anisotropy is connected to the host galaxy properties, i.e. star formation status (\S\ref{ssec:galaxy_type_anisotropy}) and host halo mass (\S\ref{ssec:angular_mass_anisotropy}). In \S\ref{ssec:model_anisotropy}, we investigate how the distribution of CGM properties varies with the implemented \simba-50 feedback models. Note that our definition of CGM is broad -- not restricted to the gas within the virial radius, but including all gas outside the virial radius that is connected to the physical activity in the central galaxies.

\subsection{Dependence on galaxy type }\label{ssec:galaxy_type_anisotropy}

\begin{figure*}
    \begin{minipage}[b]{1.0\textwidth}
        \centering
        \includegraphics[width=\linewidth]{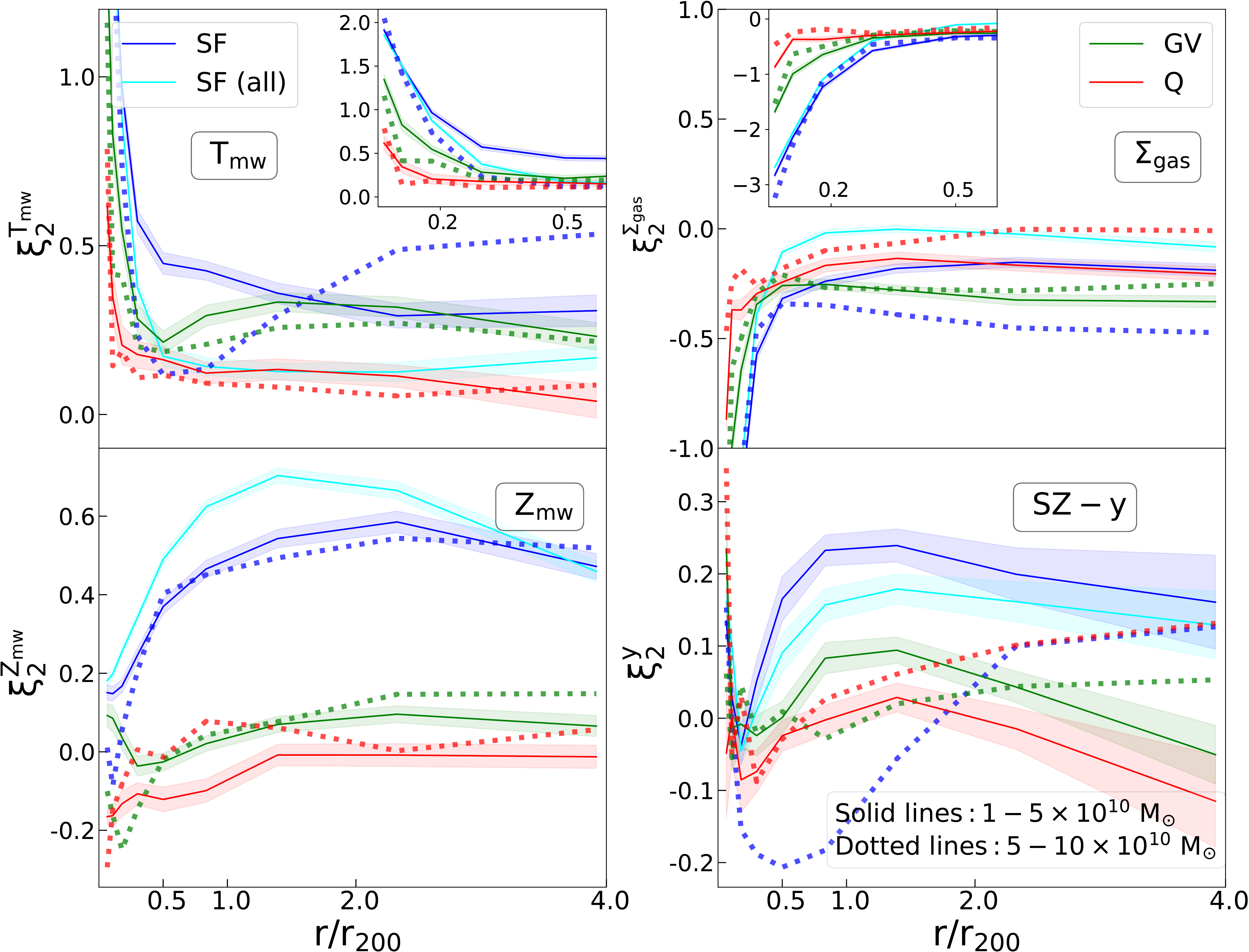}
    \end{minipage}%
    \vspace{+0.1 cm}
\caption{Comparison of mass-weighted gas temperature ($T_{\rm mw}$),  gas column density ($\Sigma_{\rm gas}$), mass-weighted gas metallicity ($Z_{\rm mw}$) and thermal SZ$-y$ ($y$) quadrupole curves for different galaxy types (SF: blue, GV: green, Q: red), as a function of projected galactocentric distance. This aims to characterise the angular dependence of the CGM properties. Results are obtained for \simba-100 `jet-active' central galaxies at $z = 0.0$ within a low stellar mass bin ($M_{\ast} = (1-5)\times10^{10}M_{\odot}$, solid lines) and intermediate mass bin ($M_{\ast} = 5-10\times10^{10}M_{\odot}$, dotted lines). For comparison, results for all SF central galaxies with accretion rate > 0.0 within the lowest stellar mass bin are shown as cyan solid lines. Shaded regions show the bootstrap errors. For clarity, we omit the error regions around the curves obtained from the intermediate mass bin. In the top row, an inset within each panel zooms into the central region with $r\lesssim 0.5r_{200c}$.}
\label{Three_type_comparison}
\end{figure*}

Generally, galaxies start out with active star formation and evolve towards a red quiescent state, passing through the GV~\citep{rodriguez_montero_2019} -- a phase where strong central suppression of star formation and lowered gas fractions are both clearly apparent \citep[see Figure \ref{central_galaxies_dist}; also ][]{Belfiore_SFMS, Sarah_simba_quenching}. AGN feedback is believed to be a major source of energy for quenching galaxies. Recently, \cite{Cui2021} showed that jet feedback is the key mechanism that quenches galaxies in the \simba\ simulations. To connect with this picture, we first examine in this section how this evolution of star forming state and stellar mass imprints itself on the CGM anisotropy. 

As discussed in \S\ref{ssec:galaxy_selection}, we subdivide the central galaxies into SF, GV and Q samples selected from the \simba-100 fiducial run, and Figure \ref{central_galaxies_dist} indicates that strong AGN feedback mostly occurs within the stellar mass range $10^{10} M_{\odot} \lesssim M_{\ast} \lesssim 10^{11} M_{\odot}$, while massive galaxies with $M_{\ast}>10^{11} M_{\odot}$ show little feedback activity. Therefore, we only discuss results obtained from galaxies within a low-mass bin (with $M_{\ast}=(1-5)\times10^{10} M_{\odot}$) and an intermediate-mass bin (with $M_{\ast}=5-10\times10^{10} M_{\odot}$). 

Figure \ref{Three_type_comparison} shows the resulting mass-weighted gas temperature ($T_{\rm mw}$), gas surface density ($\Sigma_{\rm gas}$), mass-weighted gas metallicity ($Z_{\rm mw}$), and thermal SZ$-y$ ($y$) quadrupole measurements (computed using eq. \ref{eq::quad_cal}) for \simba-100 `jet-active' central galaxies at $z = 0$ within low-mass and intermediate-mass bins. The shaded regions show the $1\sigma$ uncertainties in the quadrupole measurement determined by bootstrap resampling, as follows. For all projected galaxy maps measured from a given model, we construct a bootstrap catalogue by resampling galaxy maps with replacements but keeping the sample size the same as the original. These resampled maps are stacked as before, and the quadrupoles of each CGM property are directly measured on the normalised map. We then repeat this process 1000 times and compute the average as well as the standard deviation across the bootstrap samples. 

By inspecting the $T_{\rm mw}$ (left panel) and $\Sigma_{\rm gas}$ (second to left) curves obtained from the low-mass bin, it is clear that for SF galaxies a gaseous disk feature is prominent in the central region, causing a negative quadrupole value at $r\lesssim0.5r_{200c}$. This provides fuel to sustain star formation, although this lessens substantially going to GV and quenched systems. We will later show this quadrupolar enhancement is independent of feedback, so the anisotropy within $r\la 0.5r_{200c}$ is not a test for feedback models.

At larger radii, however, we continue to see an enhancement of gas temperatures along the minor axis (jet direction) for SF and GV populations. We will later demonstrate that this is a signature of the large-scale impact of jets on the CGM. The cumulative effect of the jet remains powerful and significant up to the limit of this plot ($\sim4r_{200c}$). For the SF case, the anisotropy falls at large distances, but it is still non-zero, indicating some residual effect from star-forming winds. For the GV case, $\xi_T$ shows a dip and then increases slightly again at $r>r_{200c}$.  

As a test, we applied the same analysis to SF galaxies without including any selection on jet activity. The resulting curves (shown as solid cyan lines) are mostly identical, except for the SF case in the low-mass bin. For these SF galaxies, the signature of the quadrupole is substantially reduced at $r>r_{200c}$, especially for the $\xi_{T}$ curve, lying instead on top of the red Q curve. This indicates that the large-scale quadrupole seen in the SF case arises from selecting only those SF galaxies with ongoing jet activity. The GV curve remains unchanged because low-mass GV galaxies already tend to be jet active in the majority of cases.

An interesting situation occurs for SF galaxies in the intermediate mass bin (dotted blue curve).  Here, there is a strong dip in $\xi_T$ between $(0.5-1)r_{200c}$, even more prominent than for GV galaxies.  This even persists when removing the active jet criterion, since at these higher masses even SF galaxies have active jets in most cases (see Figure~\ref{central_galaxies_dist}).  By visually inspecting the edge-on projected map for these systems, we notice that a cavity is present along the minor axis, and a more isotropic hot gas distribution is formed around the central region, which suggests that the signatures of inflow are being strongly curtailed relative to lower-mass SF systems while the jet activity creates significant disturbances in the CGM at larger scales.

We now consider the $Z_{\rm mw}$ results (third panel in Figure \ref{Three_type_comparison}). One can see that a significant metal enrichment at large distances is measured in the SF samples, for both stellar mass bins considered in this study. As we will demonstrate later, this is primarily the consequence of active star formation, with some enhancement from strong jet feedback pushing metal-enriched gas out to large distances. For the GV and quenched cases, metallicity anisotropy features are much weaker, since their star formation rates and hence SF-driven winds are curtailed.

We now turn to the thermal pressure anisotropy around galaxies, as would be measured by the SZ $y$ decrement ($\xi_{y}$, right panel). The level of anisotropy here is much weaker than the anisotropy seen in other thermodynamic properties. This occurs because the thermal pressure is the product of gas temperature and density, and the latter two quantities have opposite angular dependences.  This further suggests that the CGM of these galaxies rapidly equilibrate any pressure anisotropies introduced by non-spherical feedback. Unfortunately, it also suggests that SZ measures are not ideal probes of feedback-induced CGM anisotropy, unless combined with other measures that can disentangle the density and temperature.  That said, in the star forming case the opposing trends of both curves do not completely cancel out, producing a slight but significant anisotropy in the thermal pressure at $r\ga 0.5r_{200c}$.

In summary, our quadrupole analysis by galaxy SF activity indicates that at $r<0.5r_{200c}$ there are large quadrupoles in temperature and surface density that correlate with SF activity.  Beyond this, the quadrupole persists only when selecting jet-active galaxies, indicating that one must probe the small quadrupole at $r\ga r_{200c}$ in order to test AGN feedback via CGM anisotropy. Despite these features, the $y$-decrement does not show strong anisotropy, indicating that the CGM rapidly isotropises any feedback-induced pressure variations.  The metallicity anisotropy is strongest for SF galaxies, suggesting a connection to SF-driven outflows.  In \S\ref{ssec:model_anisotropy} we will flesh out these interpretations by examining how the quadrupoles depend on the physics that is included in the models. But next, we focus on exploring the mass dependence as a function of orientation angle.


\subsection{Dependence on halo mass}\label{ssec:angular_mass_anisotropy}

\begin{figure*}
    \begin{minipage}[b]{1.0\textwidth}
        \centering
        \includegraphics[width=\linewidth]{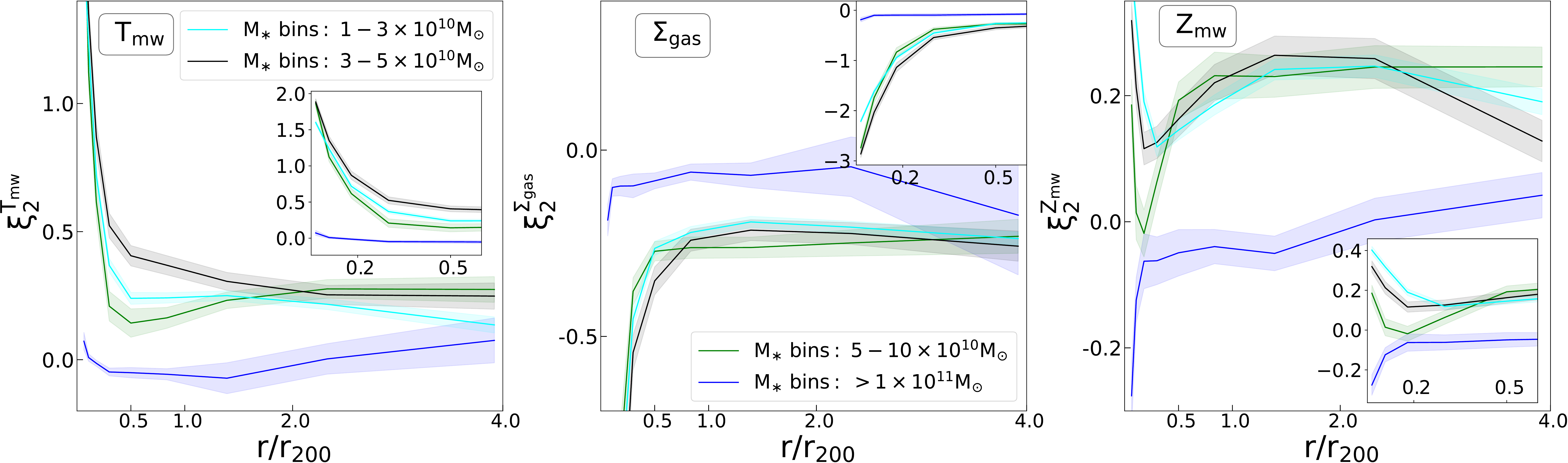}
    \end{minipage}\\
        \vspace{+0.2 cm}
    \begin{minipage}[b]{1.0\textwidth}
        \centering
        \includegraphics[width=\linewidth]{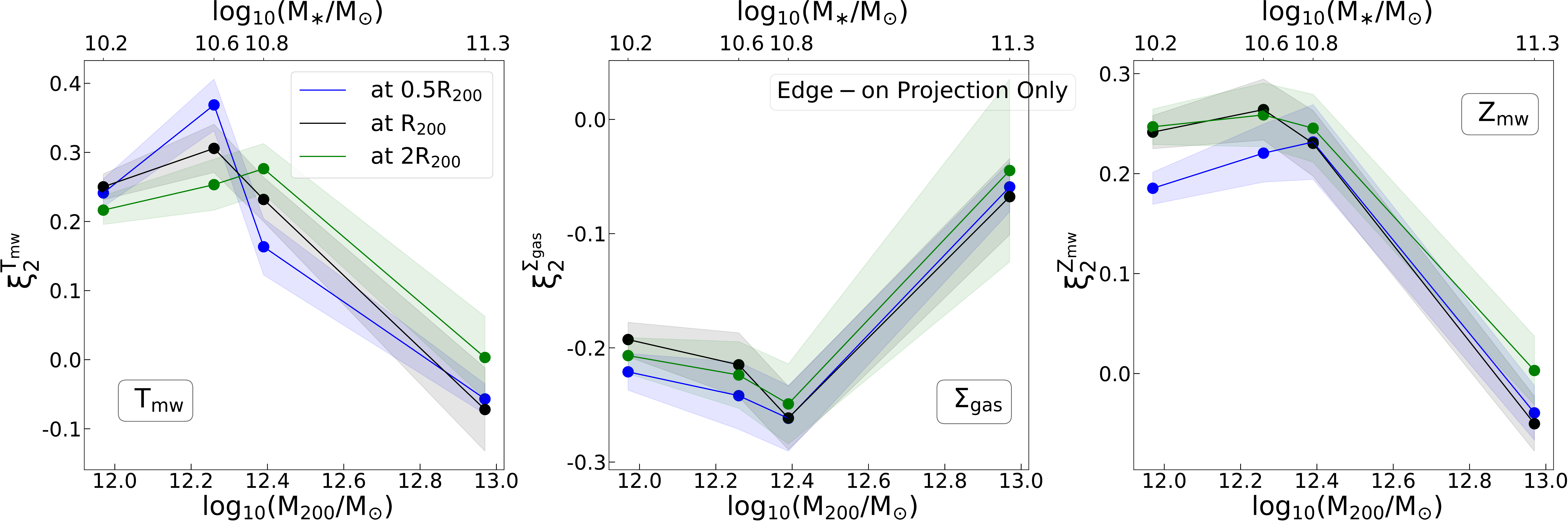}
    \end{minipage}%
    \vspace{+0.1 cm}
\caption{\textit{Upper row}: mass-weighted temperature ($T_{\rm mw}$), gas column density ($\Sigma_{\rm gas}$),and metallicity ($Z_{\rm mw}$) quadrupole curves as a function of galactocentric distance for the \simba-100 `jet-active' galaxies at $z = 0.0$. This compares the results obtained from the edge-on projected maps within four stellar mass bins. Shaded regions show the bootstrap errors.  In each panel, an inset zooms into the central region with $r\lesssim 0.5r_{200c}$.\textit{Lower row}: quadrupole values for different CGM properties evaluated at various galactocentric distances. The symbols represent the galaxy-population stacked average within a given mass bin and the shaded regions show the bootstrap errors. Only the quadrupoles measured from the edge-on view are shown here for illustration. Halo masses and stellar masses are both labelled on the same plot for comparison. For \simba galaxies, the CGM presents the most significant anisotropic features at $\textrm{log}_{10} (M_{\ast}/M_{\odot}) \sim 10.5-11.0$ or $\textrm{log}_{10} (M_{200c}/M_{\odot}) \sim 12.2-12.5$.}
\label{quadrupole_bh_sfr_across_galaxy_properties}
\end{figure*}


In the previous section we looked at the gas anisotropy for the two stellar mass bins where jet activity is seen, and focused on their connections to star forming activity. Here, we explore the trends as a function of halo mass.

The top row in Figure \ref{quadrupole_bh_sfr_across_galaxy_properties} shows the $T_{\rm mw}$, $\Sigma_{\rm gas}$ and $Z_{\rm mw}$ quadrupole radial profiles for the \simba-100 `jet-active' central galaxies within different stellar mass bins. We leave out the thermal SZ$-y$ feature due to its insignificant anisotropic levels compared to other CGM properties. Here we include all samples with stellar mass $M_{\ast}\geq 10^{10} M_{\odot}$ and then further divide them into four bins. 

The $T_{\rm mw}$ curves (upper left) show a significant change for $M_*>10^{11}M_\odot$, below which the anisotropy shows a weak dependence on halo mass, while above this mass the quadrupole essentially disappears.  More subtly, the dip in $\xi_T$ at $\sim\,$$ (0.5-1)r_{200c}$ is most prominent at $(5-10)\times 10^{10}M_\odot$, suggesting that the inflow signature is attenuating while the anisotropy induced by the cumulative effect of jets is increasing.

A similarly significant change is seen in $\xi_\Sigma$ at $M_*>10^{11}M_\odot$, above which the CGM gas suddenly becomes highly isotropic. The metallicity anisotropy shows a more complex behaviour, but again because quite low and even negative (i.e. aligned along the major axis) at high $M_*$.


As an alternative view of these trends, the bottom row of Figure \ref{quadrupole_bh_sfr_across_galaxy_properties} presents the quadrupole values evaluated at three different galactocentric distances (shown in different colours) as a function of halo mass (with the approximately equivalent stellar mass shown along the top axis), for edge-on views. The radii are chosen beyond where the influence of anisotropic accretion is dominant ($r\geq r_{200c}$), to focus on the impact of AGN jet feedback.

For the edge-on case, the anisotropies of all three quantities are maximised at similar stellar mass ranges, with  $\textrm{log}_{10} (M_{\ast}/M_{\odot}) \sim 10.5-11.0$ or $\textrm{log}_{10} (M_{200c}/M_{\odot}) \sim 12.2-12.5$. Combined with the strong amount of jet activity at lower masses ($M_*<5\times 10^{10}M_\odot$), this suggests that the net anisotropy at these larger radii is the result of the cumulative injection of bipolar feedback during the quenching process; subsequent to quenching being complete, this then rapidly becomes isotropised to leave no quadrupolar signature at the highest masses. We also performed an identical analysis for the face-on case, where the quadrupole values of all three quantities fluctuate around zero.


The `turnover' at a transitional mass probably arises from an interplay between gas content and the modes of accretion and feedback. In the \simba fiducial runs, black hole seeds are placed in galaxies with $M_{\ast} \ga 10^{9.5} M_{\odot}$. Here there is typically substantial cold gas within the inner disk region (as indicated by the $\xi_{2}^{\rm gas}$ curves), providing fuel for star formation, while the black holes accrete matter following the torque-limited model. At this stage, the black hole masses are still low: the Eddington ratios are high, so that no jet activity is triggered. With the growth of central black holes (as well as their host galaxies), the Eddington ratio drops and the accretion transitions into Bondi mode; thus the feedback becomes dominated by jets. This explains why there are no jet-driven anisotropic signals at lower mass: it takes time for the BH to accumulate enough mass to have sufficiently low accretion rates that jet feedback can be initiated.

High-mass galaxies become quenched due to the accumulated AGN feedback from their progenitors. With no gaseous disk, the jet direction becomes more stochastic, because the black hole accretion region may lack a preferred angular momentum direction. We have found that jets in \simba vary in direction on timescales of very roughly $\sim 100$~Myr (F. Jennings, priv. comm.), so this would isotropise the energy input over a long period. Furthermore, owing to the deep potential wells of such galaxies, their jets are more confined to the CGM. Those factors can all contribute towards creating a nearly isotropic CGM distribution around massive galaxies at larger scales.

\citet{TNG_CGM_anisotropy} considered the same issue using IllustrisTNG central galaxies at $z=0$. They found the mass-weighted temperature and gas density anisotropy to be peaked at $M_{\ast}\sim 10^{10.5-11} M_{\odot}$, which is consistent with our findings. However, the distance dependence of our measured CGM anisotropy is much weaker than the TNG results: they found that, for low masses, the level of anisotropy depends significantly on the radius at which it is evaluated. Also, their metallicity anisotropy falls monotonically with galaxy mass, whereas our results suggest a turnover at a transitional mass range. These differences might 
arise from different methods for estimating the anisotropy, but are more likely to reflect the different subgrid physics adopted by the two simulations.

Overall, according to the quadrupole measurements, the anisotropy of $T_{\rm mw}$, $\Sigma_{\rm gas}$ and $Z_{\rm mw}$ are all maximised around the same stellar mass range, with $\textrm{log}_{10} (M_{\ast}/M_{\odot}) \sim 10.5-11.0$ or $\textrm{log}_{10} (M_{200c}/M_{\odot}) \sim 12.2-12.5$, with the anisotropy being weaker at lower masses and completely absent at higher masses. This result does not depend on the galactocentric radius. This peak in level of anisotropy at the transitional mass probably reflects the cumulative action of bipolar jets that causes galaxy quenching, with a rapid isotropisation together with a shutting off of inflow after quenching. 

\subsection{Dependence  on \bsimba\ feedback models}\label{ssec:model_anisotropy}

\begin{figure*}
    \begin{minipage}[b]{1.0\textwidth}
        \centering
        \includegraphics[width=\linewidth]{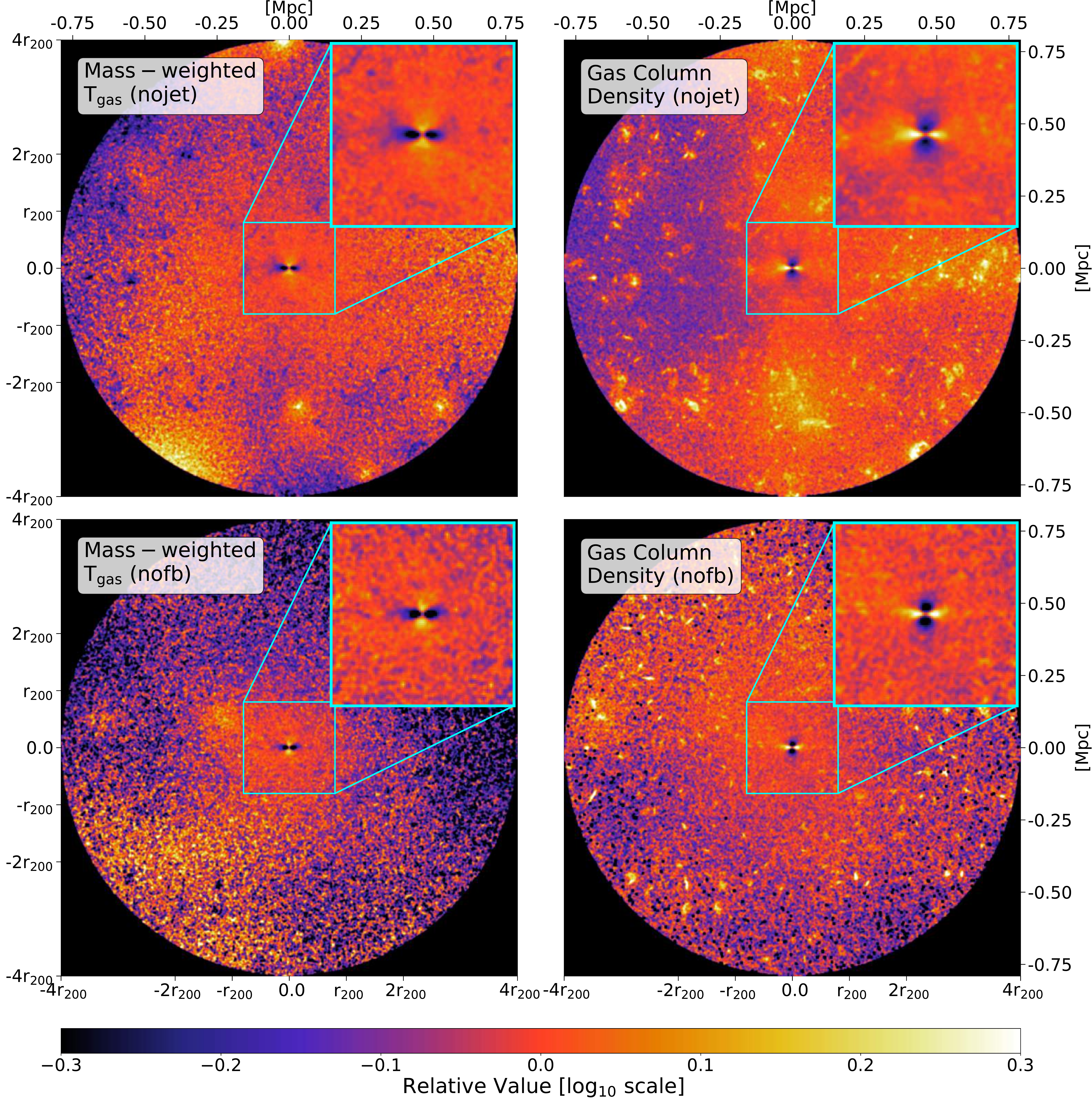}
    \end{minipage}%
    \vspace{+0.1 cm}
\caption{The same as Figure \ref{SIMBA50_allphys_bh_sfr_1e10_5e10_s151} but here showing \simba-50-`nojet' (\textit{top row}) and \simba-50-`nofb' (\textit{bottom row}) model results for comparison. Host haloes across models are matched by their halo masses. The averaged halo mass is $\langle \textrm{log}_{10} (M_{200c}) \rangle = 11.95$ for the \simba-50-`nojet' model and 11.90 for the \simba-50-`nofb' model.}
\label{SIMBA50_nojet_bh_sfr_1e10_5e10_s151}
\end{figure*}

\begin{figure*}
    \begin{minipage}[b]{1.0\textwidth}
        \centering
        \includegraphics[width=\linewidth]{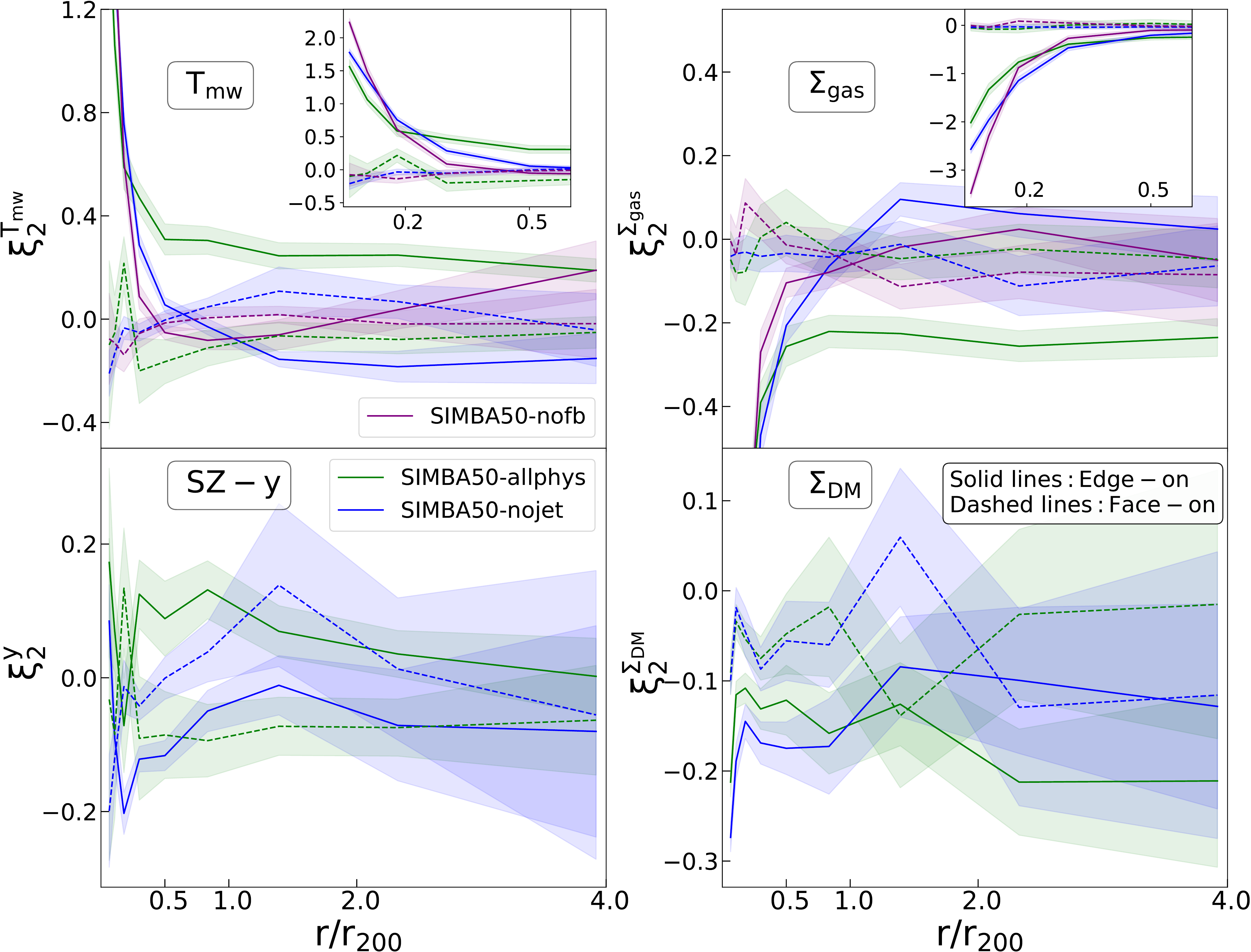}
    \end{minipage}%
    \vspace{+0.1 cm}
\caption{Quadrupole of mass-weighted temperature ($T_{\rm mw}$, \textit{top left}), gas column density map ($\Sigma_{\rm gas}$, \textit{top right}), the thermal SZ map ($y$, \textit{bottom left}) and the dark matter column density map ($\Sigma_{\rm DM}$, \textit{bottom right}), measured across \simba-50 model variants as a function of projected galactocentric distance. Results measured from both edge-on (solid lines) and face-on (dotted lines) projections are presented here. Shaded regions show the bootstrap errors. In the top row, an inset within each panel zooms into the central region with $r\lesssim 0.5r_{200c}$.}
\label{quadrupole_bh_sfr_1e10_5e10_s151}
\end{figure*}

We can most directly diagnose which feedback process is responsible for anisotropy at various scales and masses by looking at the \simba-50 variants, in which feedback mechanisms are shut off individually. 
We begin by visually inspecting the CGM distribution maps around \simba-50 galaxies evolved under different feedback models, and then quantify the differences using our quadrupole measure. 

In Figure \ref{SIMBA50_allphys_bh_sfr_1e10_5e10_s151}, we have already shown the edge-on stacked maps of CGM properties around central `jet-active' galaxies selected from \simba-50-`allphys' at $z = 0$. For comparison, we conduct the same exercise but for central galaxies selected from the `nojet'  and `nofb' run, where the former run has bipolar kinetic jet feedback plus X-ray heating turned off (top row), and the latter run has all simulated feedback activity turned off (bottom row). Host haloes across models are matched by using the dark matter particle IDs. The resulting stacked maps are showed in Figure \ref{SIMBA50_nojet_bh_sfr_1e10_5e10_s151}. In this case, we only show the stacked results for $T_{\rm mw}$ and $\Sigma_{\rm gas}$, because the anisotropic feature is not clearly seen in the $y$ and $\Sigma_{\rm DM}$ `allphys' maps. 

By inspecting the differences presented in Figure \ref{SIMBA50_allphys_bh_sfr_1e10_5e10_s151} and \ref{SIMBA50_nojet_bh_sfr_1e10_5e10_s151}, it is clear that around `allphys' galaxies, owing to the jet activity, the gas temperature is significantly higher along the minor axis (jet direction), while conversely the gas density is extended along the major axis. After turning off the jet mode feedback (top row of Figure \ref{SIMBA50_nojet_bh_sfr_1e10_5e10_s151}), the strong large-scale bipolar anisotropy in the $T_{\rm mw}$ and $\Sigma_{\rm gas}$ maps disappears. Furthermore, due to the radiative AGN winds and star-forming winds still present in the `no-jet' run, it is noticeable that the gas temperature is generally hotter and more extended around `nojet' galaxies compared to those in `nofb' (bottom row).  In contrast, the small-scale anisotropy ($\lesssim 0.5r_{200}$) in $T_{\rm mw}$ and $\Sigma_{\rm gas}$ remains present in all runs, suggesting that it is a signature of cosmological gas accretion that is independent of feedback.

Figure \ref{quadrupole_bh_sfr_1e10_5e10_s151} presents the quadrupole results of different CGM properties as a function of projected galactocentric distance, measured from different AGN models (different coloured lines), which quantifies the trends seen in the projection maps.  For reference, we also compute the quadrupole using a face-on projection (dashed lines), which as expected shows no anisotropy.

The quadrupole shows similar radial trends in the inner region ($r\la 0.5r_{200}$) among all models.  Hence the strong anisotropy within the inner halo is not a signature of feedback. To a certain degree, the inclusion of jet feedback suppresses the accretion onto the central black hole, causing a more elongated gas distribution along the major axis for models without jets.  But given the strength of the intrinsic cosmological accretion quadrupole, it will probably be difficult to test any feedback-induced anisotropy within this regime.

At larger radii, it can be seen that our quadrupole calculation is more sensitive to different feedback models for the temperature and surface density. In particular, jet feedback generates higher temperature as well as a lower density region along the minor axis, and the cumulative jet effect still remains evident at distances $\gtrsim 4r_{200c}$. These features disappear when jets are off, as showed in the `nojet' and `nofb' case. This provides a `sweet spot' for testing the AGN models in observations if using the CGM anisotropy. 


The bottom panel of Figure \ref{quadrupole_bh_sfr_1e10_5e10_s151} presents the quadrupole measurement for thermal SZ$-y$ and $\Sigma_{\rm DM}$ from the `allphys' and `nojet' models. Unlike the strongly anisotropic features visible in $T_{\rm mw}$ and $\Sigma_{\rm gas}$, here the amplitudes of both quantities fluctuate around zero. As discussed previously, the opposing trends present in the gas temperature and density cancel out the anisotropic features in the thermal pressure ($y$).  The relatively isotropic dark matter distribution (bottom right panel) on large scales indicates that the anisotropy of gas properties is not due to large-scale inflows/outflows of dynamical mass.

\begin{figure*}
    \begin{minipage}[b]{1.0\textwidth}
        \centering
        \includegraphics[width=0.8\linewidth]{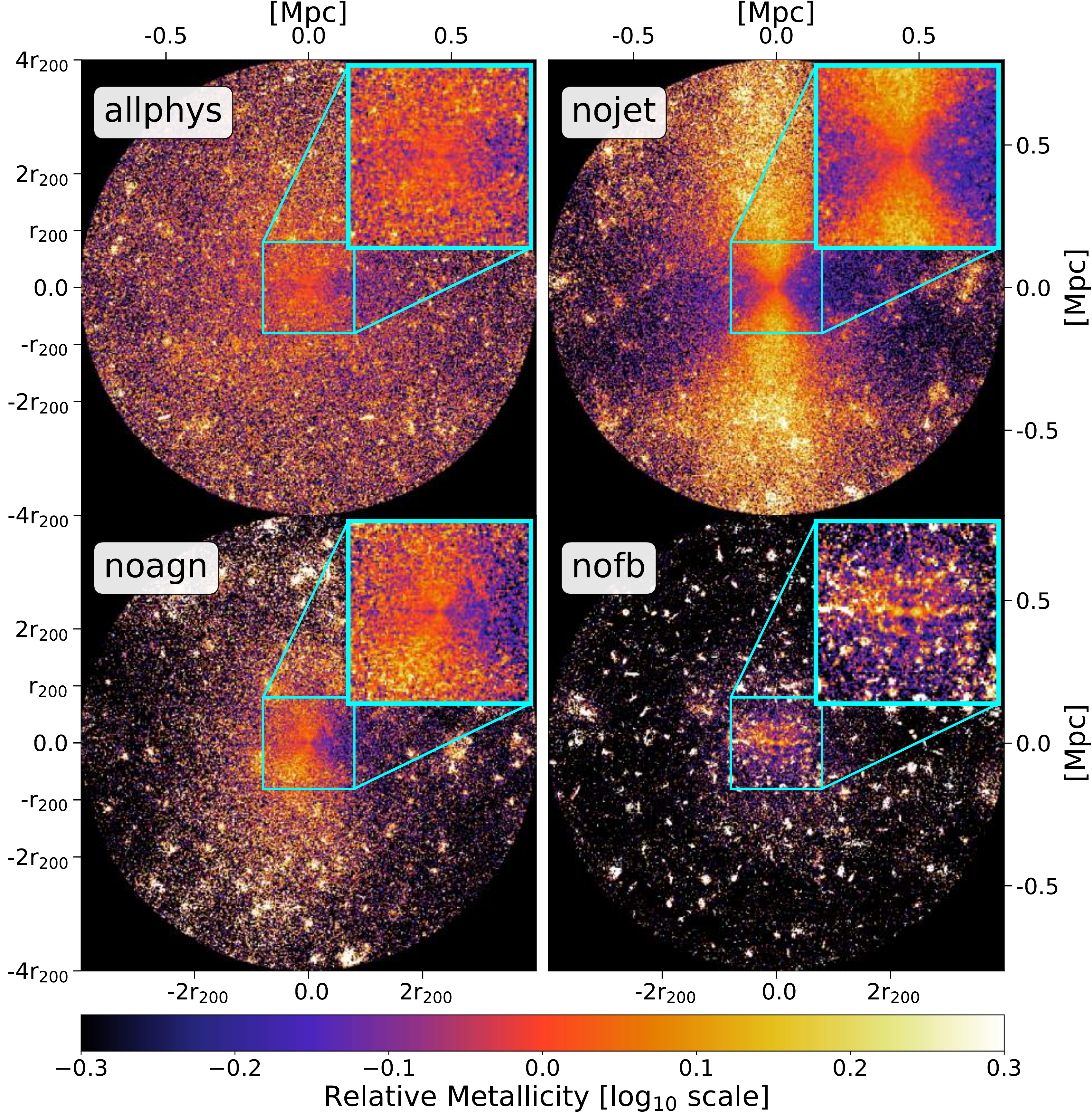}
    \end{minipage}%
    \vspace{-0.1 cm}
\caption{Anisotropy of the mass-weighted metallicity ($Z_{\rm mw}$) for `jet-active' central galaxies at $z$ = 0.0 in different \simba-50 models ($M_{\ast} = 1- 5\times10^{10} ~M_{\odot}$). Only the edge-on projection is shown. Maps are normalised with respect to the azimuthal average. Galaxies as well as their surrounding particle fields are rotated and stacked with the same approach as in Figure \ref{SIMBA50_allphys_bh_sfr_1e10_5e10_s151}. On each panel, the small window in the upper right corner zooms in to the central square region with a size of $r_{200c}$.}
\label{metal_allmodel_bh_sfr_1e10_5e10_s151}
\end{figure*}

\begin{figure}
\includegraphics[width=\columnwidth]{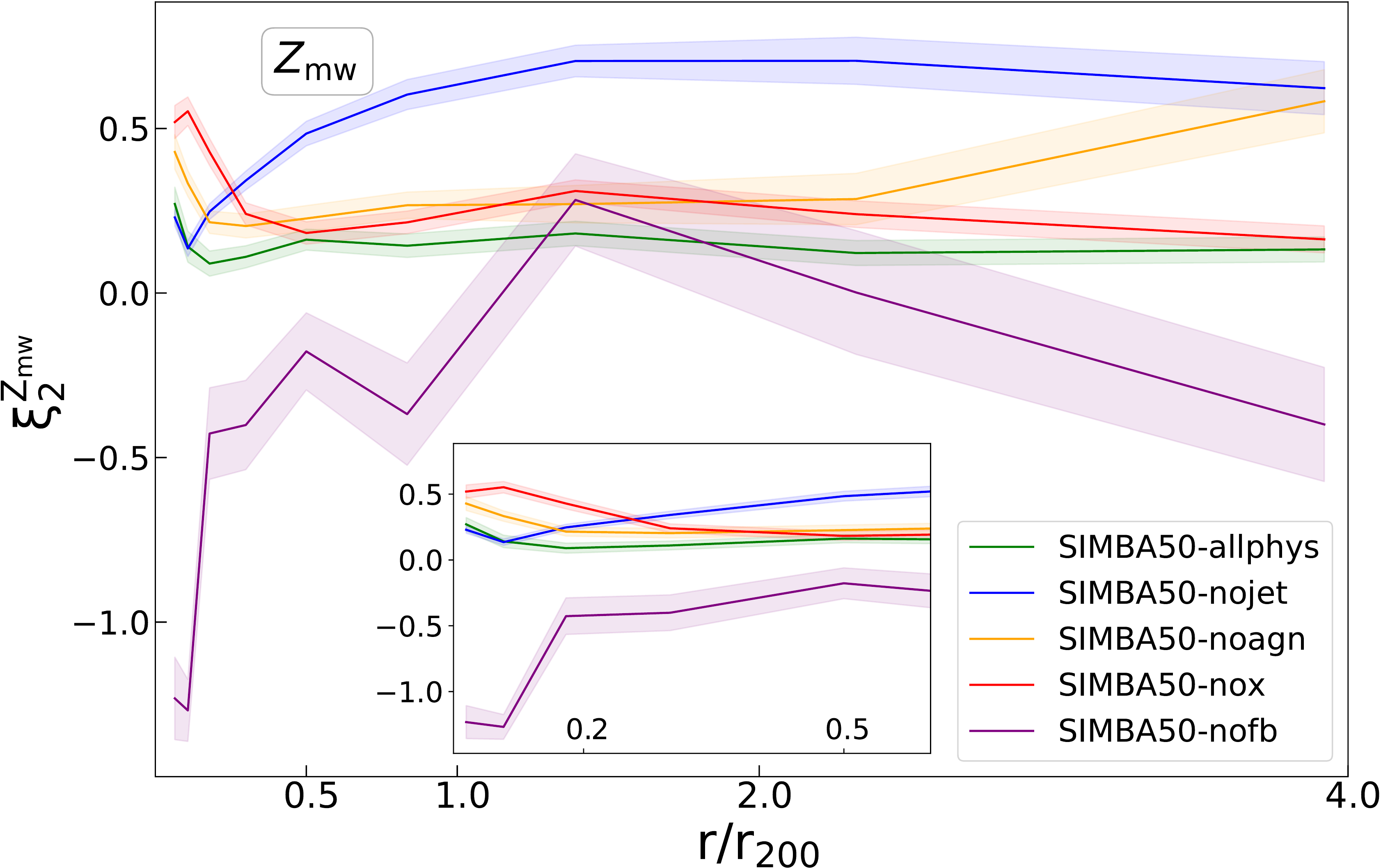}
   \caption{ $Z_{\rm mw}$ quadrupole measurement as a function of galactocentric distance in different \simba models. For clarity, only results from the edge-on projection are shown. Shaded regions show the bootstrap errors. An inset panel zooms into the central region with $r\lesssim 0.5r_{200c}$.}
   \label{quad_metal_allmodel_bh_sfr_1e10_5e10_s151}
\end{figure}

Moving on to the model dependence of $Z_{\rm mw}$ anisotropies. Figure \ref{metal_allmodel_bh_sfr_1e10_5e10_s151} shows the edge-on stacked $Z_{\rm mw}$ maps across \simba-50 models\footnote{Here we omit the presentation of the `nox' map because $Z_{\rm mw}$ anisotropies show similar trends in the `nox' and the `allphys' model.}, and Figure \ref{quad_metal_allmodel_bh_sfr_1e10_5e10_s151} shows the corresponding quadrupole measurements. We present the results obtained from all model variants because visual inspection showed that the CGM metallicity is the most sensitive property to the feedback mechanisms. This might be expected because the resulting metallicity distribution in the CGM strongly depends on both star formation status and on the gas outflow models. 

In the `allphys' model, we do not see any significant anisotropic features around the galaxies: the measured quadrupole curve is fairly flat at all distances and only slightly above zero. This is primarily due to the curtailment of star formation and SF-driven winds in central galaxies, causing a weak metal enrichment of the CGM. As can be seen from Figure \ref{Three_type_comparison}, the metallicity anisotropy can be more prominent if a sample of galaxies with high star formation rate is selected, even with active jets turned on in their hosts. The `nox' model shows similar results at larger radii, suggesting that the X-ray feedback is not the dominant factor for CGM metal anisotropy in this lowest stellar mass bin.

Interestingly, the `nojet' model shows the most significant enhancement of metals along the minor axis. Compared to the `allphys' case, the majority of `nojet' populations have strong ongoing star forming activity due to the lack of jets. Furthermore, as we shall see below (\S\ref{ssec:phase-space_diagram}, Figure \ref{phase_space_diagram_across_model_s151}), the SF wind and radiative galactic outflow can carry gas out of the central star forming regions up to $\sim\,$$ 2\,\rm cMpc$. This can effectively carry metal-rich gas out to $\sim\,$$2r_{200c}$. The `noagn' model shows weaker anisotropy because the stellar feedback is the only source that carries the metallic gas out, which is weak compared to AGN wind feedback. When all available feedback effects are turned off, metals can only be retained within galactic disks, as seen in the `nofb' model. Therefore, although the star formation status of galaxies is the primary determining factor for their $Z_{\rm mw}$ anisotropy, how far the CGM can be enriched also depends on the power of AGN winds. In \S\ref{ssec:phase-space_diagram}, we will investigate the wind power by exploring how the bulk radial gas motion depends on feedback models.

Because different studies used various tools to characterise the CGM anisotropy, it can be hard to make direct comparisons between them. Qualitatively, in agreement with the findings in the IllustrisTNG simulation \citep[e.g.][]{TNG50_galactic_outflows_driven_by_supernovae_and_black_hole_feedback,TNG_model_variants,TNG_CGM_anisotropy},  our quadrupole results suggest that the kinetic jet in \simba is powerful in reshaping the CGM properties and therefore regulating the star forming activity of galaxies. As a consequence of jet feedback, gas accretion along the major axis can be prevented in the inner region. Meanwhile, gas particles can be heated and expelled out to a large galactocentric distance along the minor axis. However, the anisotropic features in $Z_{\rm mw}$ seen in our `allphys' model are visually less prominent compared to the IllustrisTNG results \citep[e.g.][]{peroux_metal_CGM_TNG,TNG_CGM_anisotropy}, suggesting that some CGM properties are sensitive to the subgrid models adopted by different simulations. This is also apparent across \simba model variants (e.g. in Figure \ref{quadrupole_bh_sfr_1e10_5e10_s151}).

To conclude, the angular dependence of the CGM is sensitive to the feedback models. The jet activity in the centre of a galaxy regulates its CGM on large scales, at $\sim 0.5r_{200c} - 4r_{200c}$ and beyond. The cumulative jet effect causes higher temperatures and lower densities along the galaxy minor axis (jet direction), in distinction to the isotropic distribution around samples where there is no jet. Due to the anisotropic accretion, the quadrupole moments show the strongest enhancement in the inner region ($r\lesssim 0.5r_{200c}$), but this feature is independent of feedback variants and therefore cannot be used for testing the feedback models. The CGM metallicity enrichment shows an complicated interplay between star formation activity and the effectiveness of feedback-driven winds, which will be examined in the next section.


\section{feedback regulation of galaxy quenching}\label{sec:Impact_of_feedback_activity_on_galaxy_quenching}
The anisotropic distribution of the CGM presented in the previous section is a snapshot in time, but it should bear the cumulative imprint of feedback processes, and is connected to the properties of galaxies throughout cosmic time. In a way, it is a consequence of the interplay between CGM and galaxies. To understand how the the action of feedback take place, and how the anisotropy emerges throughout the evolution of the galaxies, we investigate the velocity field of the CGM in \S\ref{ssec:phase-space_diagram}, and trace the properties of progenitor galaxies in \S\ref{ssec:Redshift_evolution_of_feedback}.

\subsection{Jet activity and radial outflows}\label{ssec:phase-space_diagram}
\begin{figure*}
    \begin{minipage}[b]{1.0\textwidth}
        \centering
        \includegraphics[width=\linewidth]{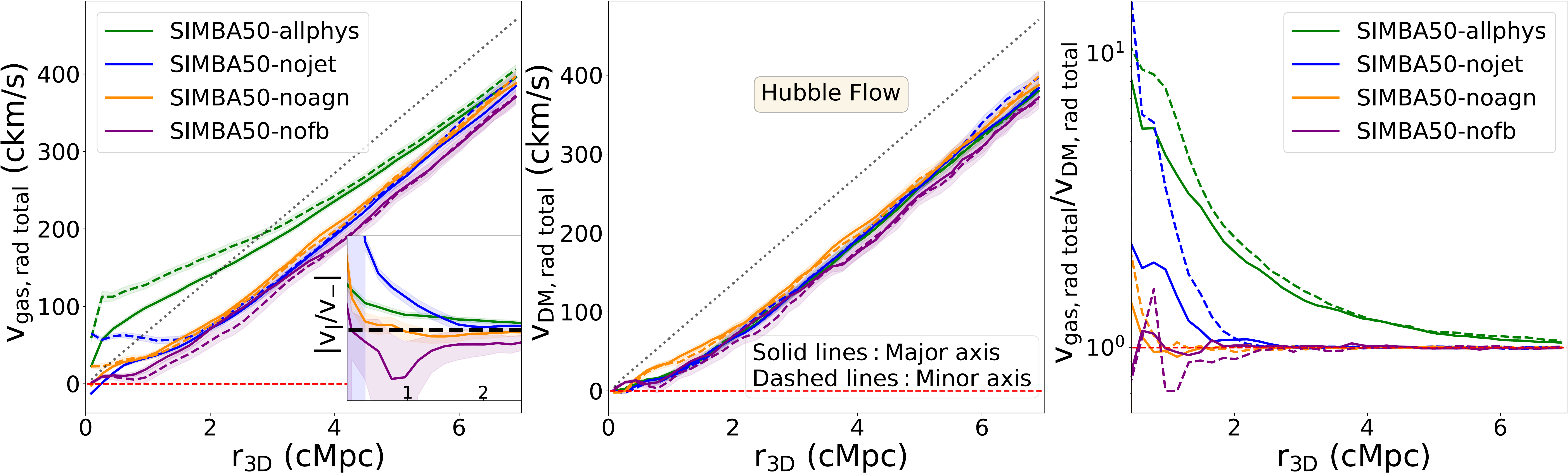}
    \end{minipage}%
    \vspace{-0.2 cm}
\caption{Total mass-weighted radial velocity for gas component (\textit{left}) and dark matter component (\textit{middle}) as a function of galactocentric distance in 3D, using `jet-active' central galaxies with $M_{\ast} = (1-5)\times10^{10} M_{\odot}$ at $z = 0.0$. Measurements are performed within a 3D cone with an opening angle of $\pm 45^{\circ}$ around the axis. Dashed lines show the results from the cone aligned with minor axis, while the solid lines show the average curves from cones along the $X$-axis and the $Y$-axis on the galaxy major plane. In the left panel, the ratio of quantities along the minor and major axes is shown as an inset in order to highlight the differences. Shaded regions represent the bootstrap errors. Models are distinguished by colours with the same colour scheme as in Figure \ref{metal_allmodel_bh_sfr_1e10_5e10_s151}. Black dotted line: Hubble flow for comparison. To highlight the effective region of each AGN variants, the ratios between the gas and dark matter curves of each model are given in the \textit{rightmost} panel for comparison.}
\label{phase_space_diagram_across_model_s151}
\end{figure*}

\begin{figure*}
    \begin{minipage}[b]{1.0\textwidth}
        \centering
        \includegraphics[width=\linewidth]{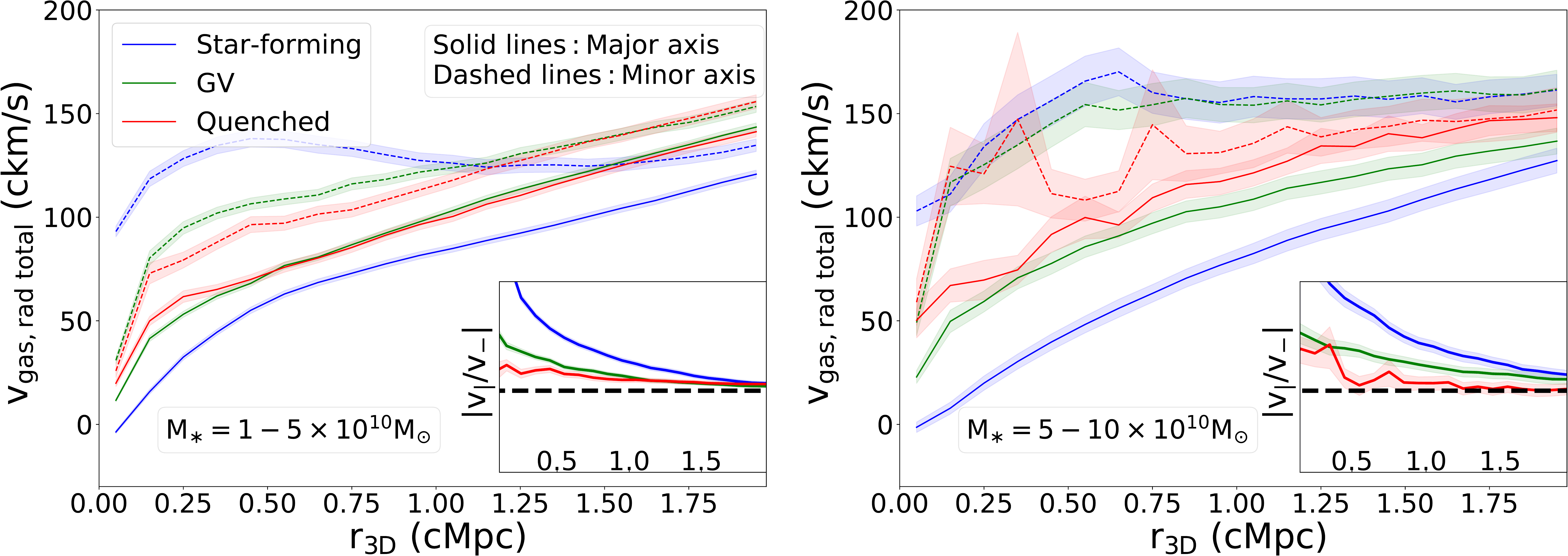}
    \end{minipage}%
    \vspace{+0.1cm}
\caption{Gas total mass-weighted radial velocity as a function of galactocentric distance in 3D, using different types of ‘jet-active’ central galaxies at $z = 0.0$ within $M_{\ast} = (1-5)\times10^{10}M_{\odot}$ (\textit{left}) and $M_{\ast} = 5-10\times10^{10}M_{\odot}$ (\textit{right}). Measurements are performed within the same cone regions as those for Figure \ref{phase_space_diagram_across_model_s151}. Shaded regions
represent the bootstrap error. Galaxy types are distinguished by colours, using the
same colour scheme as in Figure \ref{Three_type_comparison}. In both panels, the ratio of quantities along the minor and major axes is shown as an inset in order to highlight the differences.}
\label{phase_space_diagram_across_types_s151}
\end{figure*}

To see the feedback process in action, we compare the radial velocity profiles of gas along the minor and major axes. 
To make an approximate separation between the jet direction and the disk direction, we consider gas particles around each galaxy within a three-dimensional cone with an opening angle of $\pm45^{\circ}$, and measure the average radial velocity profile within each shell of the cone using the following equation: 
\begin{equation}\label{eqn::rad_vel_cal}
\overline{v}_{\textrm{\textit{par}, rad total}} (r_{\rm 3D}) = \sum_{i=0}^{N}m_{\textit{par,i}}\Bigg(\frac{ \boldsymbol{v_{i}} \cdot \boldsymbol{r_{i}}}{|\boldsymbol{r_{i}}|}+H_{0}|\boldsymbol{r_{i}}|\Bigg)\Bigg/\sum_{i=0}^{N}m_{\textit{par,i}}, 
\end{equation}
where ${\bf v_i}$ is the relative peculiar velocity of a particle;  $\bf r_i$ is its position vector relative to the black hole position; and $m_{\textit{i}}$ is the mass of a particle. The first term is the local peculiar velocity term and the second term accounts for the Hubble flow with $H_{0} = 67.74\kmsmpc$. `\textit{par}' stands for either non star-forming gas or dark matter particles. The results are shown in Figure \ref{phase_space_diagram_across_model_s151}. 

We can see that the CGM radial velocities are mostly positive (left panel), tending towards the Hubble flow at large scales, and reducing to nearly zero at small radii. This suggests that there is no gas accretion for these galaxies on average. Their host haloes have also ceased to accrete dark matter, as seen in the middle panel. We do not expect dark matter to be directly influenced by feedback processes, and thus the phase-space curves of dark matter from different models are consistent with each other. They provide references for the CGM velocities, whose ratios to the dark matter version are shown in the right-hand panel. It is clear that the CGM in the `allphys' and `nojet' models shows a stronger outflow than the dark matter version. For the `jet-active' central galaxies (green curves in the left panel), the CGM outflow is much stronger than all other models. The outflow is even stronger along the minor axis (jet direction). This is evidence that jets are responsible for these strong outflows, which induce the large-scale CGM anisotropy we have seen in the previous section.



It is worth noting that for the `nojet' model, there is also a sign of enhanced outflow within $\sim\,$$1.5\,~\rm cMpc$ along the minor axis (blue-dashed curve on the left), but this is much less effective than jet-induced AGN feedback in carrying gas out to further distances. 


These results support the physical picture that, along the minor axis, kinetic jet feedback is the most powerful driver in expelling hot gas out to large scales ($\sim\,$$3-4 \,\rm cMpc$). Additionally, radiative AGN winds are also capable of carrying gas out to $\sim\,$$1-2\,\rm cMpc$, 
but the radiative mode alone is not able to suppress the star formation activity and the chemical enrichment due to stellar feedback. This explains the strong large-scale metallicity enrichment seen for the `nojet' case in Figure \ref{metal_allmodel_bh_sfr_1e10_5e10_s151}. The effect of stellar feedback may have a minor impact within $r_{\rm 3D}\lesssim 1~\rm cMpc$, but it is much weaker compared to the AGN winds. In general, bipolar kinetic jet feedback drives gas ejection and heating out to large distances, prevents gas cooling 
in the central star forming region. We conclude that the combination of both these `ejective' and `preventative' modes is essential for the effective quenching of \simba galaxies.

When splitting the sample into Star-forming, Green Valley and Quenched populations, we find that the SF galaxies have the strongest outflows along the minor axis at small radii, followed by GV and Q galaxies (Figure \ref{phase_space_diagram_across_types_s151}). Moreover, the difference in the strength of the outflow between the minor (dashed lines) and major axes (solid lines) decreases across the three types of galaxies. This again suggests a strong correlation between jet activity and galaxy type. In addition, there may be signs of thermalisation of the energy carried out by jets. A possible scenario is that jets in SF galaxies initially carry kinetic energy out along the minor axis and the energy thermalises the CGM as they travel, increasing the CGM temperature and boosting the outflow along the minor axis. These processes gradually quench star formation, causing the galaxies to evolve from SF into GV, and eventually Q. A supporting argument for this picture is that on the left panel the blue-dashed curve is the highest at small radii ($r_{3D}<1.2~\rm cMpc$), but the green- and red-dashed curves take over at larger radii. It is possible that we are seeing the jet energy propagating outwards from the galactic centres, gradually increasing the outflow velocity at large radii while causing the galaxies to evolve to be GV and Q.


\subsection{Redshift evolution of feedback activity}\label{ssec:Redshift_evolution_of_feedback}

\begin{figure*}
    \begin{minipage}[b]{1.0\textwidth}
        \centering
     \hglue -2.0em\includegraphics[width=\linewidth]{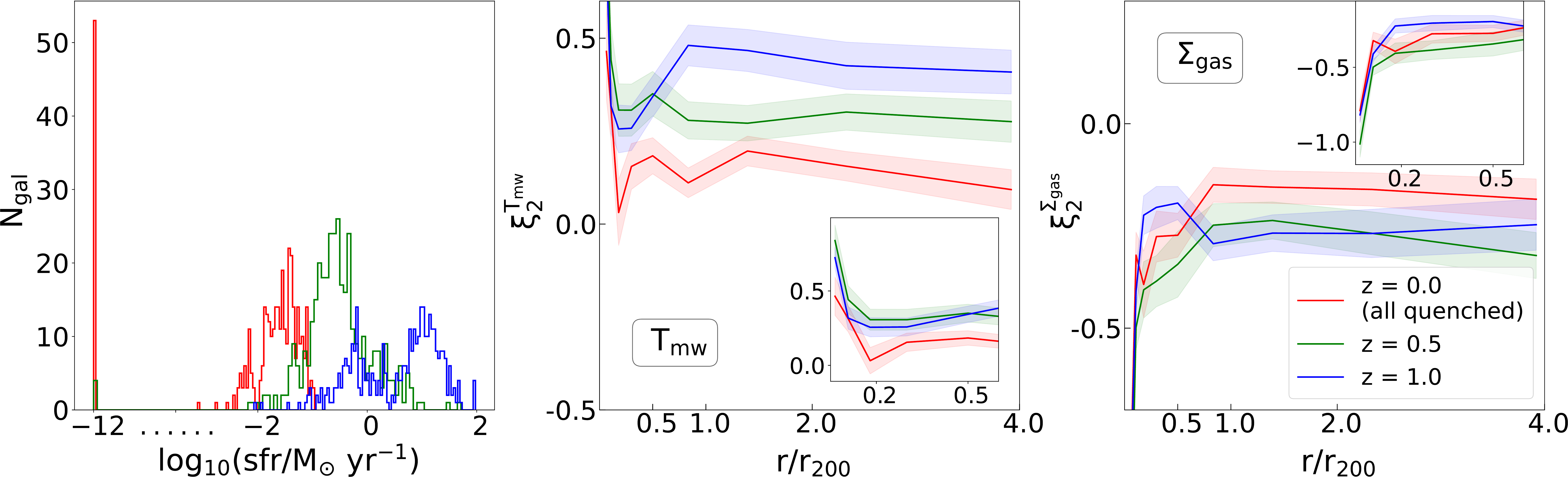}
    \end{minipage}\\
    \vspace{+0.3 cm}
    \begin{minipage}[b]{1.0\textwidth}
        \centering
    \includegraphics[width=0.97\linewidth]{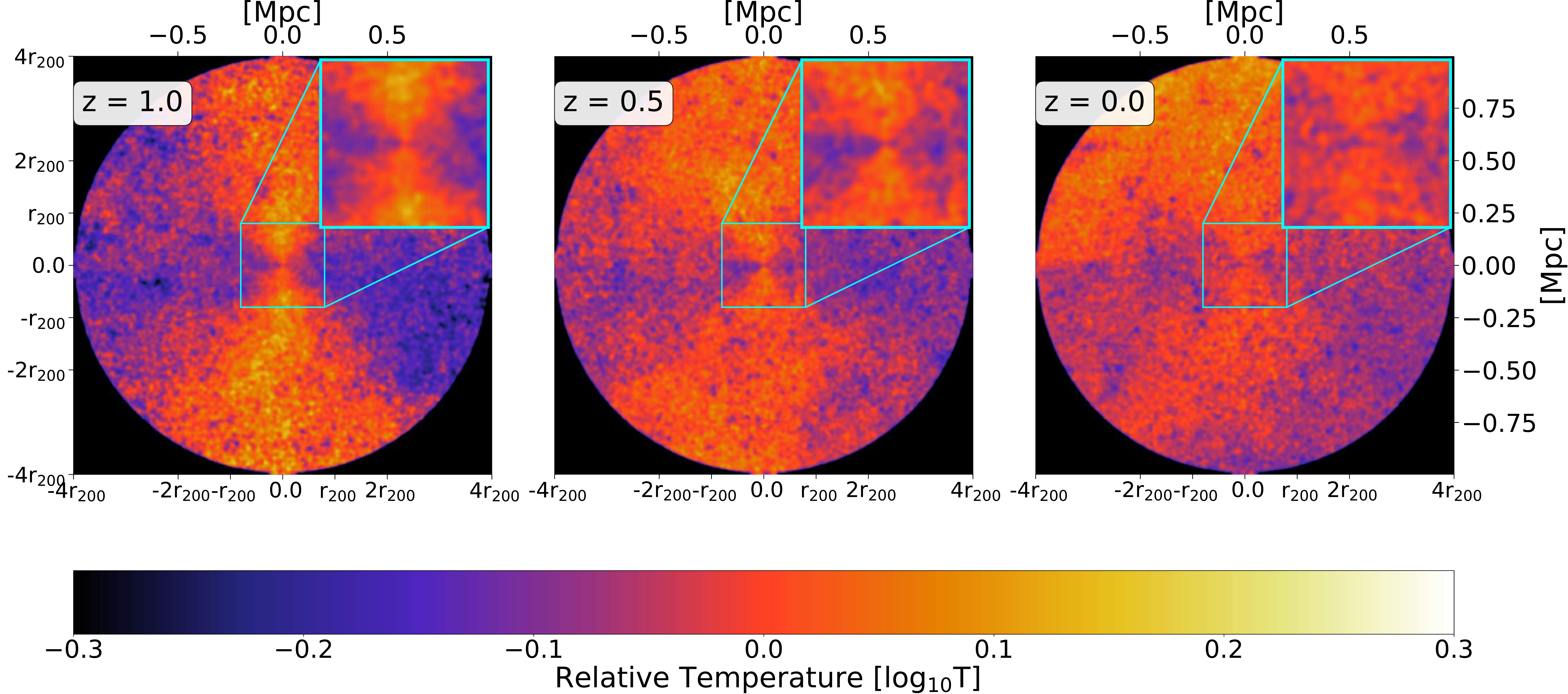}
    \end{minipage}%
\caption{Redshift evolution of feedback activity. Progenitors are traced back to $z = 1.0$ for $M_{\ast} = (1-5)\times10^{10}M_{\odot}$ quenched galaxies at $z = 0.0$, where progenitors are defined as having the most stellar particles in common. A one-to-one matching is performed within the same stellar mass range with accretion rate > 0 across snapshots. \textit{Top}: the histogram of star formation rate evolved with redshifts and the redshift evolution of $T_{\rm mw}$, $\Sigma_{\rm gas}$ quadrupole curves as a function of galactocentric distance. Only results from $z =$ 0.0 (red solid), 0.5 (green solid) and 1.0 (blue solid) are shown here for illustration. An inset within each panel zooms into the central region with $r\lesssim 0.5r_{200c}$. \textit{Bottom}: The edge-on $T_{\rm mw}$ stacked maps at different redshifts for visual inspection. It is clear that the progenitors of quenched galaxies undergo stronger AGN feedback at earlier times.}
\label{quenched_gal_zred_evl}
\end{figure*}

\begin{figure}
\includegraphics[width=\columnwidth]{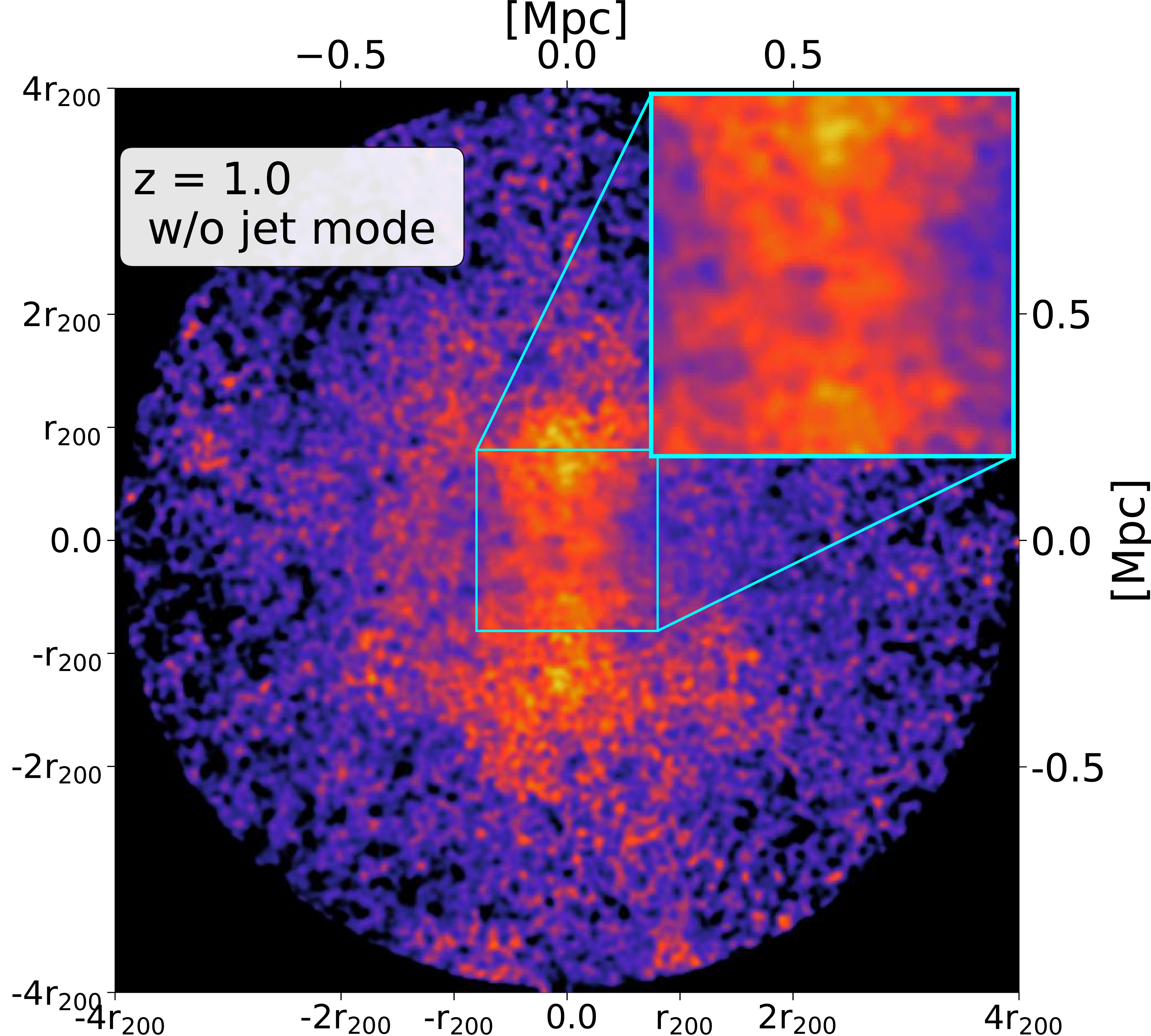}
   \caption{The edge-on $T_{\rm mw}$ stacked maps for $z = 1.0$ progenitors without kinetic-jet mode. It is clear that the star formation and radiative feedback alone are not strong enough to expel gas out to large distances.}
   \label{norm_Tmw_quenched_gal_z10_nojet}
\end{figure}

The possible scenario of galaxy evolution from SF to GV and to Q regulated by feedback processes is best examined by tracing the evolution of galaxies directly in our simulations. To do this, we identify the progenitors of a sample of quenched galaxies with $M_{\ast} = (1-5)\times10^{10}M_{\odot}$ at $z=0$. 

We identify progenitors by finding galaxies at a high redshift that have the most stellar particles in common with those at $z=0$. We make extra cuts in accretion rate and stellar mass such that galaxies at higher redshifts have accretion rate > 0 and have stellar mass within the range identical to those at $z = 0$. We do not select the progenitors to be `jet-active', in order to include the cumulative effect of all possible feedback mechanisms at higher redshifts. 
Results at $z=1$, 0.5 and $0$ are presented in Table \ref{tab:gal_type_zred_evl} and Figure \ref{quenched_gal_zred_evl}.

By default, galaxies selected at $z = 0.0$ are all `jet-active' quenched members. As shown in this table, their progenitors at $z=1$ are mainly star-forming galaxies, with a fraction of them being jet-active, i.e. they have achieved the full jet speed. Note that the majority of the remaining galaxies at $z=1$ still have jet activity, but their jets have not reached full speed. Most of these galaxies evolve into GV by $z=0.5$, and become jet-active. By $z=0$, all of them are quenched, and remain jet-active. This is also illustrated on the top-left panel of Figure \ref{quenched_gal_zred_evl}, where the continued decline of star-forming population with redshift is evident.

The $T_{\rm mw}$ and $\Sigma_{\rm gas}$ quadrupoles, measured at different redshifts, are shown in the middle and right top panel. For reference, the edge-on stacked $T_{\rm mw}$ maps at three redshifts are presented in the bottom row. Based on the $T_{\rm mw}$ quadrupole, an `ejective' effect from feedback at high redshift is clearly seen at outer region. Combining with the $T_{\rm mw}$ map given in the bottom row, one can see that those high$-z$ progenitors were undergoing strong bipolar AGN feedback at earlier times. Owing to this, gas particles were heated and ejected from the central star forming region. Hot gas then accumulated in the CGM and eventually thermalised in the outer regions, until the time when the hot gas is isotropically distributed around the central galaxies and the fuel for further star formation is depleted. Then jets are responsible for maintaining the galaxies in a quenched state at $z = 0.0$. 

At face value, there is a correlation between star-formation rate with the level of CGM anisotropy on large scales across different redshifts, i.e. as the population evolves from being star-forming at high redshift to being quenched at low redshift (Table \ref{tab:gal_type_zred_evl}), the level of CGM temperature anisotropy also decreases (middle panel of Figure \ref{quenched_gal_zred_evl}). It is not immediately clear whether it is star formation or AGN jets that are responsible for the anisotropy. To test this, we repeat our stacking analysis with a sample of galaxies at $z=1$ that have no jet activity at all, but still with star formation and radiative feedback. Results are presented in Figure \ref{norm_Tmw_quenched_gal_z10_nojet}. It is clear that the CGM for this sample is anisotropic, but the feature is present at relatively small scales, confined to be within $1-2r_{200}$. Based on this, we conclude that while other feedback processes can contribute to the CGM anisotropy on small scales, jet activity should be responsible for it on large scales. This is a distinct feature for the jet model that may provide observational signatures for detecting the direct impact of the jets on the CGM, and being able to tell the jet model apart from other feedback mechanisms.


In general, this supports the picture that strong AGN feedback plays a crucial role in effectively suppressing the star formation and maintaining the galaxies in the `quenched' state at $z \leq 1$. This once again demonstrates the strength of the \simba AGN feedback model: the feedback winds are capable of ejecting and heating up the gas to a large distance (`ejective' effect). As shown by \cite{sorini_2022}, the \simba AGN-jet feedback mode can displace baryons out to $\sim\,$$ 4\,r_{200c}$ at $z<1$ and even $\sim\,$$10\,r_{200c}$ at $z=0$, and increases the amount of hot gas in the outskirts of the halo. On relatively small scales, this process is further helped by star-formation and radiative feedback. Hot gas accumulates in the CGM and then thermalises around central galaxies, which prevents the gas from cooling and falling back onto the inner star-forming region (`preventative' effect). This gas starvation drives galaxy quenching and the jet further maintains the quenched state at low redshift. This also explains the relatively lower level of CGM anisotropy for quenched galaxies at $z=0$ shown in Figure~\ref{Three_type_comparison}.

\begin{table}
    \centering
     \caption{Redshift evolution of the galaxy types for progenitor-descendant galaxy pairs. The numbers of `jet-active' galaxies within each galaxy type are shown in the brackets}
    \begin{tabular}{|c|ccc|c|}
    \hline
         Redshift & star-forming & green valley & quenched& total\\
         \hline
         \hline
         $z=1.0$ &  218 (33) & 130 (124)&  25 (24)& 373 (181)\\
         \hline
         $z=0.5$ &  34 (27) & 249 (247)&  90 (90)& 373 (364)\\
         \hline
         $z=0.0$ &  0 (0) & 0 (0)&  373 (373)& 373 (373)\\
         \hline
    \end{tabular}
    \label{tab:gal_type_zred_evl}
\end{table}

\section{Comparisons to other simulations}\label{sec:discusion}

The anisotropic distribution of CGM is not unique to \simba. In fact other models with AGN feedback, such as the kinetic one adopted by the TNG100 simulation and the thermal one from the EAGLE simulation may also predict CGM anisotropy \citep{TNG_CGM_anisotropy, Nica_2022_x_ray_sim, Truong_2023_x_ray_anisotropy_fig}, but existing studies have not covered a range of scales as large as the one explored here (see also Figure \ref{norm_Tmw_quenched_gal_z10_nojet}).

\citet{TNG_CGM_anisotropy} carried out analyses of the TNG100 simulation that are qualitatively similar to ours; instead of using the quadrupole, the level of CGM anisotropy in their work was characterised by the minor-to-major axis ratio. The angular dependence displayed on their median stacked CGM maps, around central galaxies with $M_{\ast} = 10^{11\pm0.1} M_{\odot}$ from TNG100 at $z = 0.0$, are qualitatively similar to our results: an under-dense region of CGM gas along the minor axis with an enhancement of temperature and metallicity, even though the direction of jets is randomised at each timestep in the TNG model. In addition, the levels of anisotropy in $T_{\rm mw}$ and $\smash{\Sigma_{\rm gas}}$ are also maximised at a transitional mass range with $M_{\ast}\sim 10^{10.5-11} M_{\odot}$ ($\smash{M_{200c}}\sim 10^{12.1-12.5} M_{\odot}$). These are consistent with our findings shown in Figure \ref{quadrupole_bh_sfr_across_galaxy_properties}.

However, there are some noticeable differences between the study with TNG and our results from the `allphys' models, including (i) a more prominent metallicity ($Z_{\rm mw}$) anisotropy along the minor axis in TNG, and features in that study that monotonically decrease with galaxy masses; (ii) a stronger dependence on the galactocentric distance in TNG when evaluating the level of CGM anisotropy, especially for galaxies with $M_{\ast} \lesssim 10^{8} M_{\odot} ~(M_{\rm 200c}\lesssim 10^{12.5} M_{\odot}$); (iii) stronger anisotropic features in $T_{\rm mw}$ and $\Sigma_{\rm gas}$ for massive galaxies in TNG that are quenched and non-disky compared to their star-forming counterparts.

\citet{Nica_2022_x_ray_sim} used mock eROSITA-observations from the EAGLE simulation to study the azimuthal dependence of CGM X-ray emission. They found that the anisotropic features strongly depend on the galaxy morphology: for spheroids, the CGM distribution tends to be more isotropic, while for disk galaxies, there are clear bipolar outflows along the minor axis driven by feedback, especially seen in the stacked temperature and metallicity maps. These are consistent to our findings here in the {\sc SIMBA} model. However, their X-ray results do not show strong bi-conical features around the disc-like population. Instead, the emission is more extended along the major axes. This is because the strength of feedback-driven hot gas outflows are much weaker than the density enhancement along the discs. In terms of CGM anisotropy, \citet{TNG_CGM_anisotropy} made a comparison between the TNG and EAGLE simulations: although the two simulations predict similar CGM geometries, they produce different mass dependences: especially at low mass ($M_{200c}\sim10^{11.5}M_{\odot}$), the bipolar gas outflows around TNG galaxies are much stronger compared to those from EAGLE (see their Figure 6).

These differences in CGM properties are not surprising, given the different feedback models implemented in the simulations. As discussed in \citet{TNG_CGM_anisotropy}, the resulting CGM anisotropy around TNG, Illustris and EAGLE galaxies can be significantly affected by the detailed stellar and AGN feedback mechanisms. Also, previous tSZ$-y$ studies \citep[e.g. in][]{Tianyi_tSZ} have shown that, compared to the TNG model, the adopted \simba feedback model is more energetic at heating and expelling gas into the CGM. Therefore, gas in the CGM can be thermalised more efficiently around galaxies, producing a more isotropic hot gas distribution around massive quenched galaxies compared to those in the TNG simulation. This could explain why the anisotropy levels around \simba quenched galaxies are lower that those seen in TNG. Meanwhile, the strong suppression of star formation owing to the powerful feedback causes the depletion of metal-enriched gas in the central region. As seen in Figure \ref{Three_type_comparison}, the level of $Z_{\rm mw}$ anisotropy can be much more prominent compared to the GV and Q counterparts. Beyond that, it is worth nothing that the CGM anisotropic signals along the major axes can also be driven by non-feedback inflows, as discussed by some simulation studies \citep[for further details, see e.g.][]{Hafen_2022_FIRE,Stern_2023}.

\section{Implications for observations}\label{sec:discussion_2}
We have used the \simba model to show that jet activity from AGN can provide powerful energy feedback, which regulates star-forming activity for the galaxy, and leaves imprints on the properties of the CGM. This opens possibilities for constraining feedback models with observational signatures of the CGM. It has been shown that the CGM properties around galaxies, such as the azimuthal asymmetry and the X-ray projected profile, have a large dependence on the adopted feedback models. Based on the mock observations from TNG, EAGLE and \simba simulations, recent studies have shown the potential ability of detecting these differences in more detail using future X-ray microcalorimeters \citep{Ayromlou_paper,Schellenberger_2023,Truong_2023_x_ray_anisotropy_fig}, e.g. from Line Emission Mapper \citep[LEM,][]{LEM}. In fact, by stacking galaxies from the state-of-the-art X-ray observations \citep[e.g. from eROSITA:][]{erosita_ref}, the CGM properties, such as the projected luminosity and radial profiles have already presented clear differences around different types of galaxies \citep{comparat_2022,Chadayammuri_2022}. These observations, however, show some discrepancies with simulated results, which may indicate imperfections in the current feedback models. 

In terms of the CGM anisotropy, one key step in being able to identify the direction of the jet for the stacking of the CGM, which can be challenging in practice. However, one can use other observable means to infer the jet directions. The stellar disk is an observable that may correlate with the jet direction due to the expected alignment of the angular momentum of the centre of a disk galaxy with that of the stellar disk, or even with the HI gas. We have explicitly studied the angular momentum of the BH and stellar disk from the \simba simulations, and confirmed that there is a strong, though not perfect, correlation between them. This is further evident from the results presented in Figure \ref{simba50_simba100_convergence_test}, where we repeat our analyses for the anisotropy of the CGM, but instead of using the jet directions, we use the stellar angular momentum vectors as proxies. We see that the results remains similar. Therefore, at least in \simba simulations, the direction of the stellar angular momentum is a good approximation for the jet direction. This provides one possible guide for future observational analyses of this kind.  

From both observations and simulations, the satellite quenching fraction is found to be anisotropically distributed around the central hosts, where a larger quenching fraction is found along the minor axis compared to that within the major plane \citep[e.g. in][]{Zaritsky_gal_anisotropy, quenched_fraction_anisotropy}. Also, previous studies have shown that the spin vectors of high-mass haloes and galaxies (with $M_{\rm halo}\gtrsim10^{12} h^{-1} M_{\odot}$) are on average perpendicular to the filament axes \citep[e.g. in][]{halo_fil_alignment_mil,halo_spin_fil_eagle}. This provides us with some possible indirect means of linking the direction of galactic outflow with large-scale accretion flows. 

Once the jet direction is found, the large-scale CGM anisotropy in temperature and gas density is a clear signature that can be sought in observational data. It is also clearly distinct from the relatively small-scale features induced by other feedback processes \citep{TNG_CGM_anisotropy}. We have shown that the thermal SZ effect, which is sensitive to the gas pressure, is probably not a good observable for this because the anisotropy shown in the SZ-$y$ maps is weak due to the cancelling effect between an increased gas temperature and a decreased gas density. However, the X-ray emission is sensitive to the square of the gas density, and only weakly on the gas temperature. This is different from that of the tSZ signature. The combination of X-ray data with tSZ measurements may allow us to measure the temperature map of the CGM \citep[e.g.][]{Adam2017}.

Furthermore, the most anisotropic direction or jet direction can be directly determined using observational data, such as from X-ray and radio surveys. The current state-of-the-art X-ray missions, \textit{Chandra} and \textit{XMM-Newton} \citep{ XMM_ref, chandra_ref}, have already revealed giant cavities and hot gas shock fronts along the minor axes of massive galaxies \citep[e.g.][]{Hlavacek-Larrondo_X_ray, Liu_X_ray}. Specifically, new generation X-ray surveys such as \textit{eROSITA} are capable of sampling millions of AGN by scanning the whole sky within a much wider range of energies \citep[0.2-10 keV:][]{erosita_ref}. \citet{TNG_CGM_anisotropy} have demonstrated that the predicted X-ray hardness from an \textit{eROSITA}-like survey can be helpful and promising for capturing  bipolar CGM features, and this property is quite sensitive to the adopted feedback models in simulations. However, one needs to stack a large number of samples ($\gtrsim10^{4}$) to reach a signal-to-noise ratio greater than 3 around halo mass $M_{200c}\sim10^{12} M_{\odot}$. The all-sky coverage of \textit{eROSITA} allows the stacking of large samples and allows us to conduct anisotropy analyses of the thermodynamic gas properties in the CGM. For AGN possessing collimated radio jets, telescopes such as the \textit{VLA} and \textit{LOFAR} have good sensitivity to extended sources \citep{VLA_ref, LOFAR_Croston}. As suggested by our results, the accumulated CGM anisotropy features from powerful jet feedback should still be significant up to several Mpc. The advent of these surveys will enable the investigation of feedback-induced CGM anisotropy out to larger radii.

\section{Conclusions and Discussion}\label{sec:conclusions}

We have used \simba simulation model variants to explore how the properties of galaxies and their circumgalactic medium (CGM) are regulated by feedback processes, with a focus on the jet activity of their central black holes. We find that at redshift $z<1$, central galaxies with active AGN jet feedback are most commonly found in the stellar mass range between $(1-5)\times10^{10}M_{\odot}$, and most commonly within the green valley (Figure \ref{central_galaxies_dist}). Driven by the powerful bipolar jet-activity, the CGM becomes anisotropic in temperature, gas density and metallicity at the galactocentric distance of $4r_{200}$ and beyond. This is supported by direct evidence from the outflow of the gas extended to several cMpc, and that the outflow is stronger along the jet direction. We have also show evidence that jet activity is responsible for driving the galaxy to evolve from being star-forming to occupying the green valley, and eventually moving into a quenched state. 

On large scales ($r> 0.5r_{200}$), the jets increase the gas temperature around the minor axis of a galaxy while suppressing its density. These features are similar to those seen in star-formation and radiative feedback processes, but the latter occurs at relatively small scales ($r<2 r_{200}$). The difference in the scale of impact on the CGM is potentially a unique signature for distinguishing the jet model from other type of feedback processes.
Due to the cancellation effect, the resultant gas pressure remains relatively isotropic. This makes it challenging to use the thermal SZ effect alone to detect the effect of the jets on the CGM, but it may be possible with the combination of SZ and X-ray observables. 

The CGM metallicity is strongly enhanced along the minor axis around SF galaxies, peaking at around $r_{200}$. This disappears for GV galaxies, and Q galaxies even show a slight metallicity enhancement along the major axis. Given that star formation is the expected source for metal enrichment, it is likely that star formation provides the metals, which are then carried out by the jets to larger scales. 
We do not find any obvious anisotropy for the dark matter, as expected. 

On small scales ($r< 0.5r_{200}$) we also find strong anisotropy for the CGM, but this is common to different feedback models, even when all explicit feedback is turned off, indicating that it is associated with anisotropic cosmological accretion rather than any feedback process.

Understanding how feedback activity regulates the baryonic cycle is a crucial step in galaxy formation and evolution. From the above analysis, it is apparent that the CGM anisotropic features depend strongly on the feedback models, and it is a consequence of both `ejective' and `preventative' effects that leave an imprint on the resulting CGM properties. The \simba `allphys' simulation and its model variants provide us an ideal opportunity to compare and predict what may happen around galaxies selected from different environments. By observing the CGM distribution, one can further infer the host galaxy properties, such as the star formation status and the impact of feedback. 

While this study has focused on theoretical aspects, we hope to provide inspiration and observational directions for future CGM studies, such as from which direction one expects to observe the strongest CGM anisotropy signal. Apart from the central region, the accumulated effect of feedback activity around low$-z$ galaxies can be `observed' out to several Mpc, so there are good prospects of making the necessary measurements in practice. With the advent of new surveys and improved simulation models, these probes of anisotropic CGM properties should enable us to learn ever more about the detailed operation of feedback in galaxy formation.

\section*{Acknowledgements}
We are grateful for the publicly available simulations from the \simba\ project. 
During this work, DS and JAP were supported by the STFC consolidated grant no. RA5496. DS was further supported by the Swiss National Science Foundation (SNSF) Professorship grant no. 202671.
 WC is supported by the STFC AGP Grant ST/V000594/1 and the Atracci\'{o}n de Talento Contract no. 2020-T1/TIC-19882 granted by the Comunidad de Madrid in Spain. He also thanks the Ministerio de Ciencia e Innovación (Spain) for financial support under Project grant PID2021-122603NB-C21 and ERC: HORIZON-TMA-MSCA-SE for supporting the LACEGAL-III project with grant number 101086388.
RD acknowledges support from the Wolfson Research Merit Award program of the U.K. Royal Society. YC acknowledges
the support of the Royal Society through a University Research Fellowship and an Enhancement Award.
This work used the DiRAC@Durham facility managed by the Institute for Computational Cosmology on behalf of the STFC DiRAC HPC Facility. The equipment was funded by BEIS capital funding via STFC capital grants ST/P002293/1, ST/R002371/1 and ST/S002502/1, Durham University and STFC operations grant ST/R000832/1. DiRAC is part of the National e-Infrastructure.

\section*{Data availability}
The raw \simba simulation data and halo catalogues used in this paper are available at {\tt https://simba.roe.ac.uk}.
The remaining data will be made available on request to the lead author.

\bibliographystyle{mnras}
\bibliography{draft}
\appendix

\section{Numerical convergence}\label{sec::convergence_test_full_physics_models}

\begin{figure*}
\centering
    \includegraphics[width=1.0\linewidth]{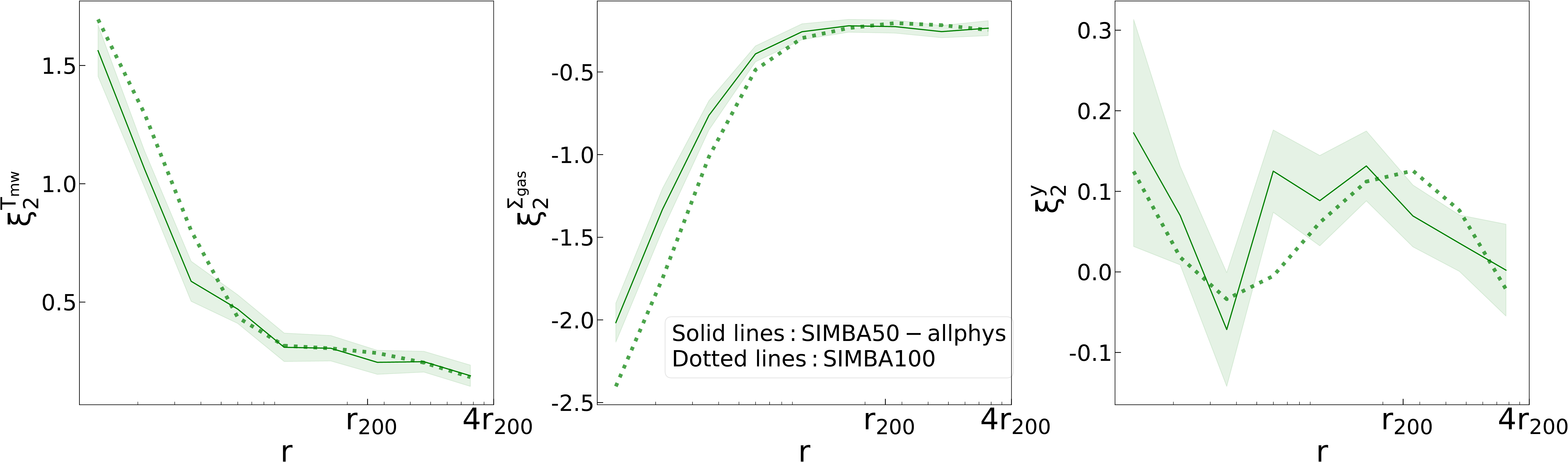}
\caption{$T_{\rm mw}$, $\Sigma_{\rm gas}$ and thermal SZ$-y$ quadrupole curves measured around samples from the \simba-50-`allphys' (solid lines)  and the \simba 100 model (dotted lines), using all `jet-active' galaxies with stellar masses between $1\times10^{10}M_{\odot}$ to $5\times10^{10}M_{\odot}$ at $z = 0.0$. Only results from the edge-on projected maps are shown here. The shaded area shows the bootstrap errors around the average, where only error bars obtained from the \simba-50-`allphys' case are included.}
\label{simba50_simba100_convergence_test}
\end{figure*}

For the \simba-50-`allphys' model and the \simba -100 model, both runs implement the same feedback mechanisms and have identical numerical resolutions. Therefore, we want to see the sensitivity of the predicted CGM anisotropy to the finite simulation volume by testing whether the quadrupole curves measured from two `allphys' models converge well.

Figure \ref{simba50_simba100_convergence_test} shows he resulting quadrupole curves around galaxies selected from the \simba-50-`allphys' model and the \simba -100 model, using all `jet-active' galaxies with stellar masses between $1\times10^{10}M_{\odot}$ to $5\times10^{10}M_{\odot}$ at $z = 0.0$. Only results from the stacked edge-on $T_{\rm mw}$, $\Sigma_{\rm gas}$ and thermal SZ$-y$ maps are shown here for illustration. The shaded area shows the bootstrap errors around the average, where only error bars obtained from the the \simba-50-`allphys' are included. According to this plot, the general trends of CGM anisotropic features measured from two models are in good agreement with each other. Especially at $r\gtrsim 0.5 r_{200c}$ -- the region where the cumulative effect of the jet is the most noticeable, the quadrupole values estimated from two models overlap reasonably well. Hence, we conclude that the resulting features of CGM anisotropy are not sensitive to the finite volume size.

\section{Effect of stacking axes}\label{sec::convergence_test_stacking}

In this study, in order to exhibit the angular dependence of CGM properties, galaxies as well as their surrounding particle fields are stacked with respect to the minor axes of their inner gas discs, or `jet-direction' around `allphys' galaxies. However, this vector direction is hard to measure in real observations. An alternative is to choose the stellar angular momentum vector, which is computed using all stellar particles within the stellar half-mass radius around the galactic centre. In this section, we examine the sensitivity of the predicted CGM anisotropy to the reference axes that we stack against.

\begin{figure*}
    \begin{minipage}[b]{0.97\textwidth}
        \centering
        \includegraphics[width=0.97\linewidth]{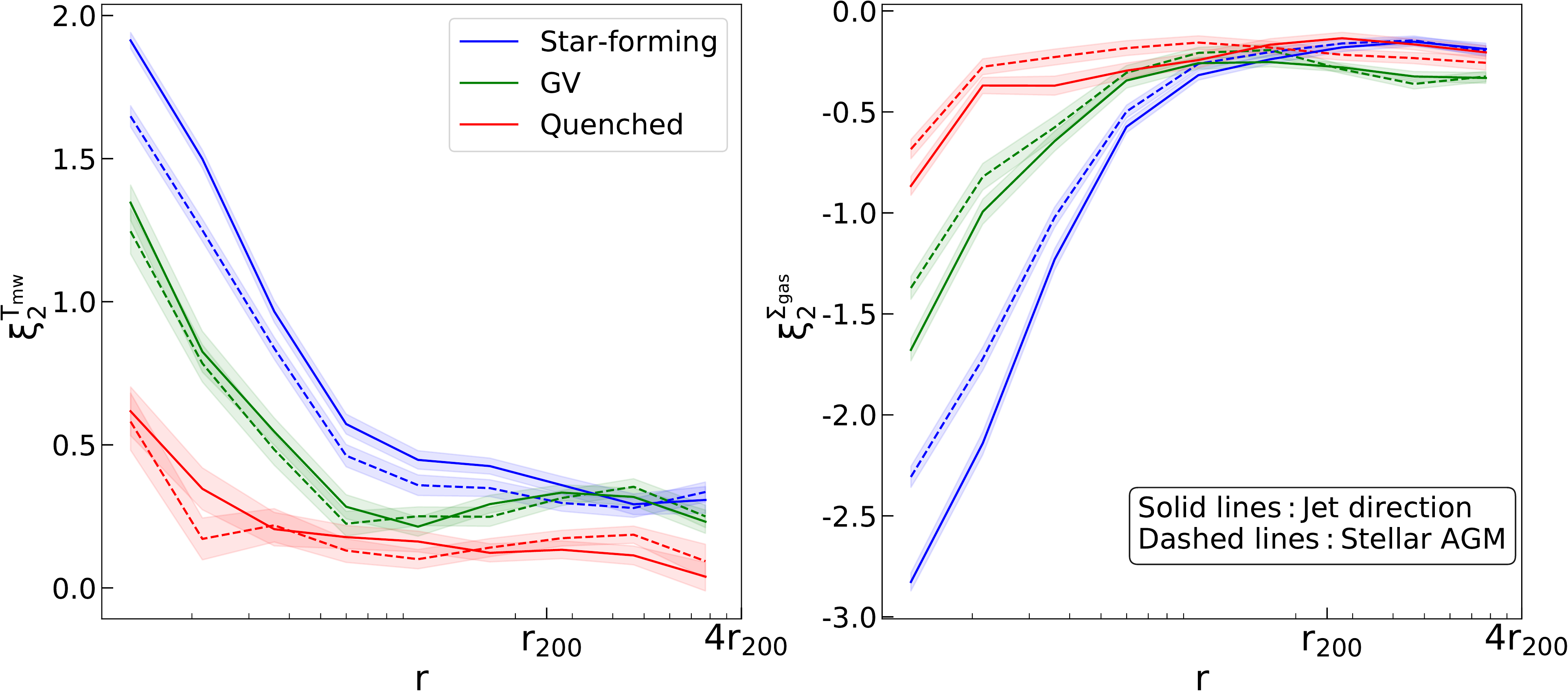}
    \end{minipage}\\
    \vspace{-0.3 cm}
    \begin{minipage}[b]{0.97\textwidth}
        \centering
        \includegraphics[width=0.97\linewidth]{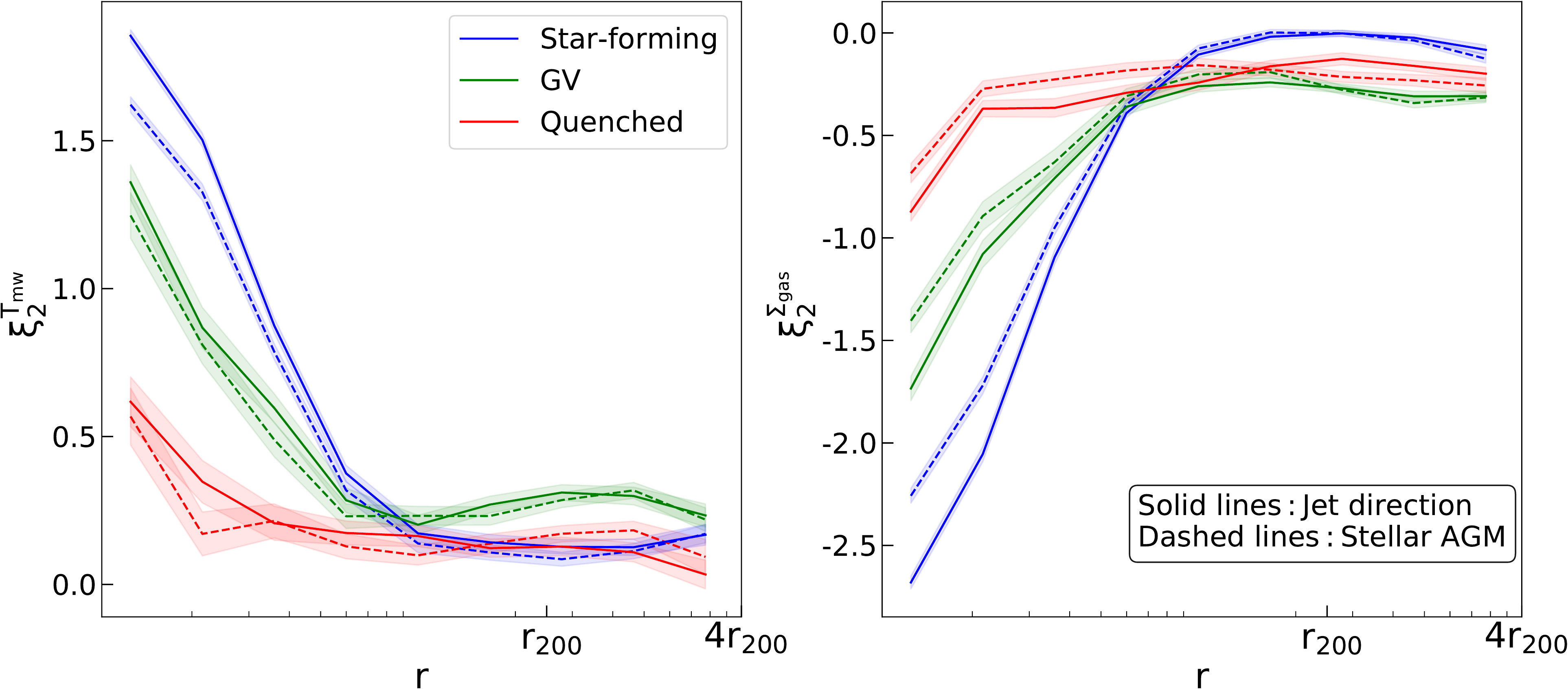}
    \end{minipage}%
    \vspace{-0.05 cm}
\caption{Comparison of $T_{\rm mw}$ and $\Sigma_{\rm gas}$ quadrupole curves for different galaxy types from the \simba-100 model (SF: blue, GV: green, Q: red) measured on the edge-on projected maps stacked along the jet direction (as shown in Figure \ref{Three_type_comparison}, solid lines) and along the stellar angular momentum vector, dashed line. Top row shows the results measured around all `jet-active' central galaxies within stellar mass bin $M_{\ast}=(1-5)\times10^{10}M_{\odot}$, and the bottom row shows the results measured around all central galaxies with accretion rate $>\,$0.0 within the same stellar mass bin. Shaded regions show the bootstrap errors.}
\label{stellar_comp_1e10_5e10}
\end{figure*}

Figure \ref{stellar_comp_1e10_5e10} shows the comparison of $T_{\rm mw}$ and $\Sigma_{\rm gas}$ quadrupole curves for galaxies from the \simba-100 model, measured on the edge-on projected maps stacked along the jet direction or along the stellar angular momentum vectors. Here, for illustration, we test the sensitivity of the resulting CGM anisotropy to the reference stacked axes around galaxies with different star formation status. The top row of Figure \ref{stellar_comp_1e10_5e10}, shows the results measured around `jet-active' \simba-100 central galaxies at $z = 0.0$ within the stellar mass bin $M_{\ast}=(1-5)\times10^{10}M_{\odot}$. For comparison, the bottom row shows the results measured around all central galaxies with accretion rate $>\,$0.0 within the identical stellar mass bin. 

According to the figure, the general quadrupole trends and CGM features, especially at large distances, are in agreement with each other -- regardless of stacking the galaxies along their jet or stellar angular momentum vectors. The two vectors for the star forming and the GV populations align reasonably well, with an average angle between them of $\sim\,$$37^{\circ}$ for SF populations and $\sim\,$$60^{\circ}$ for GV populations. For quenched populations, however, the average angle between the two vectors can reach $\sim\,$$80^{\circ}$. Around quenched galaxies, it is highly possible that well-formed gaseous and stellar disks are missing, so that their vector directions can have a more uniform distribution compared to their SF and GV counterparts. As shown in \S\ref{ssec:galaxy_type_anisotropy}, the cumulative jet effect drives a more extended and isotropic hot gas distribution around the quenched galaxies, which makes them less sensitive to the choices of stacking axes in the inner region. Therefore, we conclude that the resulting CGM properties and their angular dependence in the stacked maps are insensitive to those two chosen axes. We leave a more thorough comparison with other choices of axes for future work.

\label{lastpage}
\end{document}